\documentclass[pra,twocolumn,floatfix,superscriptaddress,longbibliography, nofootinbib]{revtex4-1}
\usepackage{amsmath,amsfonts,amssymb,amsthm,graphics,graphicx,epsfig,bbm}
\usepackage[colorlinks=true,citecolor=blue,linkcolor=blue,urlcolor=blue]{hyperref}
\usepackage[usenames]{color}
\usepackage[dvipsnames]{xcolor}
\usepackage{float}
\usepackage{algorithm}
\usepackage{algorithmicx}
\usepackage{algpseudocode}
\usepackage{graphicx}
\usepackage{tikz}
\usepackage{subfigure}
\usepackage{amsmath}
\usepackage{epsfig}
\usepackage{dcolumn}
\usepackage{bm}
\usepackage{color}
\usepackage{times}
\usepackage{epstopdf}
\usepackage{amscd}
\usepackage{amsthm}
\usepackage{amssymb}
\usepackage{bbm}
\usepackage{amssymb}
\usepackage{amstext}
\usepackage{latexsym}
\usepackage{comment}
\usepackage{hyperref}
\usepackage{amsfonts}
\usepackage{psfrag}
\usepackage{soul,xcolor}
\usepackage[normalem]{ulem}
\usepackage{dsfont}
\usepackage{txfonts}
\usepackage{physics}
\usepackage{footnote}
\usepackage{multirow}
\usepackage{appendix}
\usepackage{mathtools}
\usepackage{tikz}
\usetikzlibrary{quantikz}
\usepackage{bbold}
\usepackage{enumitem}
\usepackage{xstring}
\usepackage{cancel}
\usepackage{amscd}
\usepackage{bbm}
\usepackage{units}
\usepackage{cancel}
\usepackage{xstring}
\usepackage{import}
\usepackage{subfiles}
\usepackage{amsmath,amsfonts,amssymb,amsthm,graphics,graphicx,epsfig,bbm}
\usepackage[colorlinks=true,citecolor=blue,linkcolor=blue,urlcolor=blue]{hyperref}
\usepackage[usenames]{color}
\usepackage{graphicx}
\usepackage{subfigure}
\usepackage{amsmath}
\usepackage{epsfig}
\usepackage{dcolumn}
\usepackage{bm}
\usepackage{color}
\usepackage{times}
\usepackage{epstopdf}
\usepackage{amssymb}
\usepackage{amstext}
\usepackage{latexsym}
\usepackage{hyperref}
\usepackage{amsfonts}
\usepackage{psfrag}
\usepackage{soul,xcolor}
\usepackage[normalem]{ulem}
\usepackage{dsfont}
\usepackage{txfonts}
\usepackage{physics}
\usepackage{xcolor}
\usepackage{footnote}
\usepackage{multirow}
\usepackage{appendix}
\usepackage{mathtools}
\usepackage{ulem}
\usepackage{mathtools}

\definecolor{mycolor}{RGB}{106,81,162}

\usepackage{mathtools}

\newcommand{\iden}[1]{
    \ifthenelse{\equal{1}{\string #1}}
  {
   \mathbbm{1}
  }
  {
   \mathbbm{1}^{\otimes#1}}
  }
\newcommand{\ketzero}[1]{
    \ifthenelse{\equal{1}{\string #1}}
  {
   \ket{0}
  }
  {
   \ket{0}^{\otimes#1}}
  }
\newcommand{\brazero}[1]{
    \ifthenelse{\equal{1}{\string #1}}
  {
   \bra{0}
  }
  {
   \bra{0}^{\otimes#1}}
  }
\newcommand{\ketone}[1]{
      \ifthenelse{\equal{1}{\string #1}}
    {
     \ket{1}
    }
    {
     \ket{1}^{\otimes#1}}
    }
  \newcommand{\braone}[1]{
      \ifthenelse{\equal{1}{\string #1}}
    {
     \bra{1}
    }
    {
     \bra{1}^{\otimes#1}}
    }

\usepackage{todonotes}

\usepackage{physics}  
\usepackage{dsfont}
\usepackage{amssymb}
\usepackage{txfonts,color}
\usepackage{multirow}
\usepackage{booktabs} 
\makeatother

\begin{document}

\title{Quantum approximated cloning-assisted density matrix exponentiation}

\author{Pablo Rodriguez-Grasa}
\email[Corresponding author: ]{\qquad pablojesus.rodriguez@ehu.eus}
\affiliation{Department of Physical Chemistry, University of the Basque Country UPV/EHU, Apartado 644, 48080 Bilbao, Spain}
\affiliation{EHU Quantum Center, University of the Basque Country UPV/EHU, Apartado 644, 48080 Bilbao, Spain}
\affiliation{TECNALIA, Basque Research and Technology Alliance (BRTA), 48160 Derio, Spain}

\author{Ruben Ibarrondo}

\affiliation{Department of Physical Chemistry, University of the Basque Country UPV/EHU, Apartado 644, 48080 Bilbao, Spain}
\affiliation{EHU Quantum Center, University of the Basque Country UPV/EHU, Apartado 644, 48080 Bilbao, Spain}

\author{Javier Gonzalez-Conde}

\affiliation{Department of Physical Chemistry, University of the Basque Country UPV/EHU, Apartado 644, 48080 Bilbao, Spain}
\affiliation{EHU Quantum Center, University of the Basque Country UPV/EHU, Apartado 644, 48080 Bilbao, Spain}

\author{Yue Ban}
\affiliation{Instituto de Ciencia de Materiales de Madrid (CSIC), Cantoblanco, E-28049 Madrid, Spain}


\author{Patrick Rebentrost}

\affiliation{Centre for Quantum Technologies, National University of Singapore, Singapore 117543}

\author{Mikel Sanz}
\email[Corresponding author: ]{\qquad mikel.sanz@ehu.eus}
\affiliation{Department of Physical Chemistry, University of the Basque Country UPV/EHU, Apartado 644, 48080 Bilbao, Spain}
\affiliation{EHU Quantum Center, University of the Basque Country UPV/EHU, Apartado 644, 48080 Bilbao, Spain}
\affiliation{IKERBASQUE, Basque Foundation for Science, Plaza Euskadi 5, 48009, Bilbao, Spain}
\affiliation{Basque Center for Applied Mathematics (BCAM), Alameda de Mazarredo, 14, 48009 Bilbao, Spain}

\date{\today}

\begin{abstract}

    Classical information loading is an essential task for many processing quantum algorithms, constituting a cornerstone in the field of quantum machine learning. In particular, the embedding techniques based on Hamiltonian simulation techniques enable the loading of matrices into quantum computers. A representative example of these methods is the Lloyd-Mohseni-Rebentrost protocol, which efficiently implements matrix exponentiation when multiple copies of a quantum state are available. However, this is a quite ideal set up, and in a realistic scenario, the copies are limited and the non-cloning theorem prevents from producing more exact copies in order to increase the accuracy of the protocol. Here, we propose a method to circumvent this limitation by introducing imperfect quantum copies, which significantly improve the performance of the LMR when the eigenvectors are known. 

\end{abstract}

\maketitle

\section{Introduction}
The successful implementation of many quantum algorithms relies heavily on the efficient encoding of classical data into quantum computers. In this sense, an inefficient encoding can undermine any potential quantum advantage offered by the processing stage, specially for those implementations comprising quantum machine learning techniques \cite{HHL, HHL_Childs,Preconditioned_HHL, Data_Fitting,  Biamonte_2017, Patrick_SVM_2014,  lloyd2013quantum, Robert_power_data, Schuld_2019, Patrick_PCA,Encoding_variational, lloyd2020quantum}. In addition to this, there is no universal loading protocol, and each specific case must be carefully studied to adapt the encoding to the particular problem being addressed \cite{Encoding_variational, Schuld2021, lloyd2020quantum}. These encodings and their applications range from loading probability distributions into the amplitudes of quantum states as a means to implement a Monte Carlo integration or as an initial condition of a differential equation \cite{Patrick_Options, marinsanchez2021quantum,gonzalezconde2023efficient, gonzalezconde2022simulating}, to calculate the exponentiation of a certain matrix in order to calculate its principal components \cite{Patrick_PCA, Martin_2021} or to solve an associated linear system of equations \cite{HHL, HHL_Childs, Preconditioned_HHL, martin2022digitalanalog}.  In this sense, density matrix exponentiation constitutes a particular instance of the Hamiltonian simulation problem \cite{Kimmel_2017, Trotter_1959, Suzuki_1976, Barends_2016, Layden2022, Childs_2021, Campbell_2019, Low_2019,Berry_2015, Childs2012,Berry_2014,Berry_2015_2, Efekan_2022, Low_2017, C_rstoiu_2020, Dong_2021}, in the specific scenario where the Hamiltonian matches a density matrix. This technique also represents a case of special relevance among the methodologies used to accelerate the application of certain unitaries with precomputation, particularly for tasks like fidelity estimation \cite{gilyén2022improved, marvian2016universal}, opening the door to the possibility of achieving speedups ranging from quadratic to exponential \cite{huggins2023accelerating}.

To achieve Hamiltonian exponentiation, typical methods often rely on various assumptions, such as the sparsity of the Hamiltonian \cite{Berry_2006} or the availability of multiple copies of a quantum state \cite{Patrick_PCA}. In the latter case, the celebrated Lloyd-Mohseni-Rebentrost (LMR) protocol was proposed as an efficient method for density matrix exponentiation (implementing the unitary $e^{-i\rho t}$) \cite{Patrick_PCA, Kimmel_2017}, and it has even been experimentally realized \cite{LMR_experiment}. This protocol has proven to be a crucial subroutine in several quantum machine learning applications, including the construction of quantum Support Vector Machines \cite{Patrick_SVM_2014}, Bayesian Deep Learning on quantum computers \cite{Zhao_2019}, the Helstrøm quantum classifier \cite{lloyd2020quantum}, quantum linear regression \cite{Schuld_regression}, quantum discriminant analysis \cite{discriminant}, and quantum capsule networks \cite{quantum_capsule}.

Nonetheless, the main drawback of these scenarios is that, in general, we do not have an unlimited number of copies available. Additionally, given an arbitrary unknown quantum state whose preparation is not possible, the no-cloning theorem \cite{no_clonning} states that it is forbidden to generate independent and identical copies of this state. To address this limitation, the concept of imperfect cloning using quantum cloning machines was introduced. In this context, one of the most relevant proposals in the literature is the universal quantum cloning machine \cite{PhysRevA.54.1844, Wang_2011} (UQCM), which provides a state-independent optimal result for a single qubit, generating a copy with theoretical fidelity up to 5/6 with respect to the original state. However, this fidelity scales very poorly with both, the number of copies and the dimension, limiting the applicability of this cloning machine. Alternatively, the biomimetic cloning of quantum observables \cite{Alvarez-Rodriguez_2014} (BCQO) offers a state-dependent result, potentially generating copies with better fidelity in exchange for the requirement to select a preferred basis for cloning. Although the copies generated are not perfect, the quantum information transmitted in this process may be sufficient to enhance quantum algorithms that require density matrix exponentiation or are based on the availability of multiples copies of a quantum state
as the non-linear transformations of density matrices \cite{Zoe_holmes, Kimmel_2017}, the Virtual Distillation for Quantum Error Mitigation \cite{Huggins_2021} or when learning from experiments \cite{Huang_2022}. 

In this article, we propose an enhanced LMR protocol by combining the original methodology with biomimetic copies of a quantum state. Under certain assumptions, this combined methodology enables us to improve the performance of the LMR protocol when the number of copies is scarce or limited. By considering $n$ original copies, we generate $k$ bio-mimetic copies from each one. The introduction of these copies leads to a dramatic reduction in the number of required original copies to achieve the same accuracy levels when the size of the system increases significantly. Our arguments offer valuable insights and intuition, glimpsing that the average enhancement in the statistical case scales with the dimension of the quantum system. Consequently, when the generation of the quantum state $\rho$ comes from an experiment, an expensive computation, or is simply unknown, the introduction of biomimetic copies is postulated as an advantageous approach.
\\

\section{Preliminaries}
In the following section, we provide a detailed overview of the LMR protocol and the concept of biomimetic cloning, as these are the foundational elements necessary for understanding our contribution.
\subsection{The LMR protocol} 
\begin{figure}[t!]
\centering
\includegraphics[width=0.9\columnwidth]{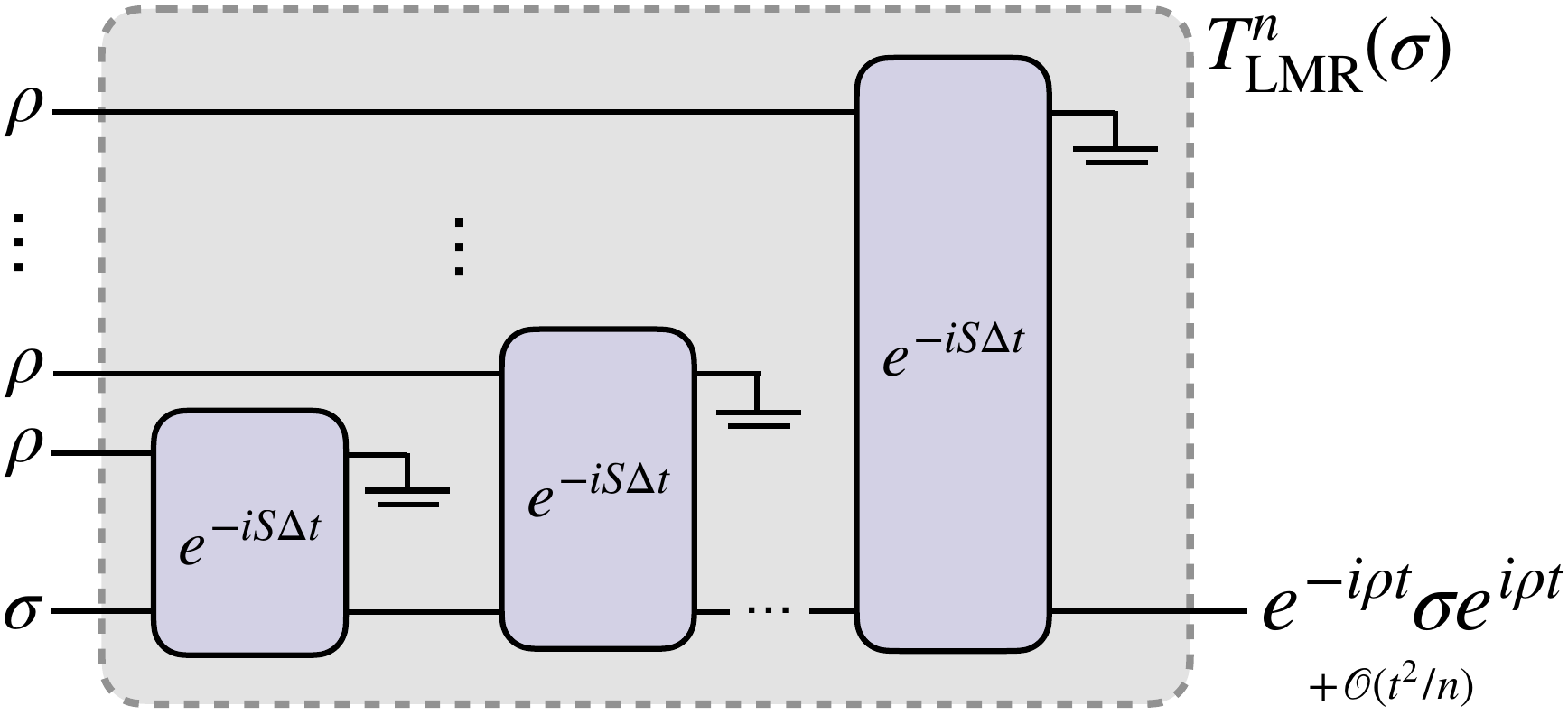}
\caption{Description of the implementation of the LMR protocol using $n$ copies of $\rho$. The quantum channel $T_\mathrm{LMR}^{n}$ acting on $\sigma$ approximates the target operation $e^{-i\rho t}\: \sigma \: e^{i\rho t}$, with an error on the order of $\mathcal{O}(t^2/n)$.}
\label{fig:LMR}
\end{figure}

Let us consider that we have a quantum state $\rho$ generated by the oracle $\hat{O}_s$. The LMR protocol strives to implement the complex exponentiation of such quantum state, acting on a quantum state $\sigma$, by assuming the availability of $n$ copies of  $\rho$ \cite{Patrick_PCA}. Each step of this methodology uses a single copy of  $\rho$ and implements an operation on the state of the system $\sigma$ described by the quantum channel
\begin{equation}
    \begin{split}
        T_\mathrm{LMR}(\sigma)&=\tr_2[e^{-i\Delta t S} (\sigma \otimes \rho)\ e^{i \Delta t S} ]\\
        &=\cos^2\Delta t \: \sigma+ \sin^2\Delta t \: \rho - i \: \sin\Delta t \: \cos\Delta t \: [\rho,\sigma].
    \end{split}
    \label{eq:LMR}
\end{equation}
Therefore, after $n$ iterations the state of the system results into
\begin{equation}
    T_\mathrm{LMR}^{n}(\sigma) = \tr_{1^{\perp}}\left[\prod_{j=2}^{n+1} e^{-i\Delta t S_{1,j}} \sigma \otimes \rho^{\otimes n} \left(\prod_{j=2}^{n+1} e^{-i\Delta t S_{1,j}}\right)^{\dag} \right],
    \label{eq:output_LMR}
\end{equation}
where $\tr_{1^{\perp}}$ denotes the partial trace on all registers except the first one, $S_{ij}$ denotes the swap operation between registers $i$ and $j$. This process is illustrated in Fig.~\ref{fig:LMR}. Expanding this expression we get
\begin{equation}\label{ntimes_channel}
\begin{split}
    T_\mathrm{LMR}^{n}(\sigma)=&\sum_{k=0}^n \binom{n}{k} (-i \sin{\Delta t})^k\;\cos^{2n-k}\Delta t \: [\rho,\sigma]_k \\
    &+ \rho\;(1-\cos^{2n}{\Delta t})
\end{split}
\end{equation}
where $[\rho,\sigma]_{k+1}=[\rho,[\rho,\sigma]_k]$ is the nested commutator and $[\rho,\sigma]_{k=0}=\sigma$. Considering up to the $\Delta t^2$ contributions
\begin{equation}
\begin{split}
    T_\mathrm{LMR}^{n}(\sigma)=&\sigma-i[\rho,\sigma]\:n\:\Delta t-\frac{1}{2}[\rho,\sigma]_2 \:n(n-1)\: \Delta t^2\\
    &+(\rho-\sigma)\:n\:\Delta t^2+\mathcal{O}\bigl(\Delta t^3\bigr).
\end{split}
\end{equation}
On the other hand, using the Baker-Campbell-Hausdorff lemma, we can develop the target operation
\begin{equation}\label{Baker}
    e^{-i\rho t}\:\sigma\:e^{i\rho t}=\sum_{k=0}^\infty \frac{1}{k!} [\:-i\rho t,\:\sigma\:]_k=\sigma-i\:[\rho,\sigma]\:t-\frac{1}{2}[\rho,\sigma]_2\:t^2+\mathcal{O}\bigl(t^3\bigr).
\end{equation}
For the purpose of equalising the $n\Delta t$ contributions we choose $n=t/\Delta t$. To compare these two last equations and obtain the error made using the LMR protocol, $\text{$\varepsilon$}_{\mathrm{LMR}(n)}$, we calculate the norm between the two outputs
\begin{equation}
    \begin{split}
        \text{$\varepsilon$}_{\mathrm{LMR}(n)}= \norm{T_\mathrm{LMR}^{n}(\sigma)-e^{-i\rho t}\:\sigma\:e^{i\rho t}}_1   \\ = \frac{t^2}{2n}\big|\big|\;[\rho,\;\sigma]_2+2(\rho-\sigma)\ \big|\big|_1
        &+\mathcal{O}(t^3/n^2)\leq \mathcal{O}(t^2/n).
    \end{split}
    \label{error_LMR}
\end{equation}
where $\norm{A}_1=\mathrm{tr}\sqrt{AA^\dagger}$ denotes the norm-1 or trace norm. Hence, to simulate $e^{-i\rho t}\:\sigma\:e^{i\rho t}$ up to an accuracy $\text{$\varepsilon$}$,  $n\sim~\mathcal{O}(t^2/\text{$\varepsilon$})$ original copies of $\rho$ become needed. In \cite{Kimmel_2017}, this protocol was demonstrated to be optimal in terms of the number of copies needed in the asymptotic limit. More recently, in \cite{bound_dme}, it was shown that, in the non-asymptotic regime, the sample complexity of the LMR protocol is upper bounded by $4t^2/\epsilon$.

\begin{figure*}[t!]
\centering
\includegraphics[width=.92\textwidth]{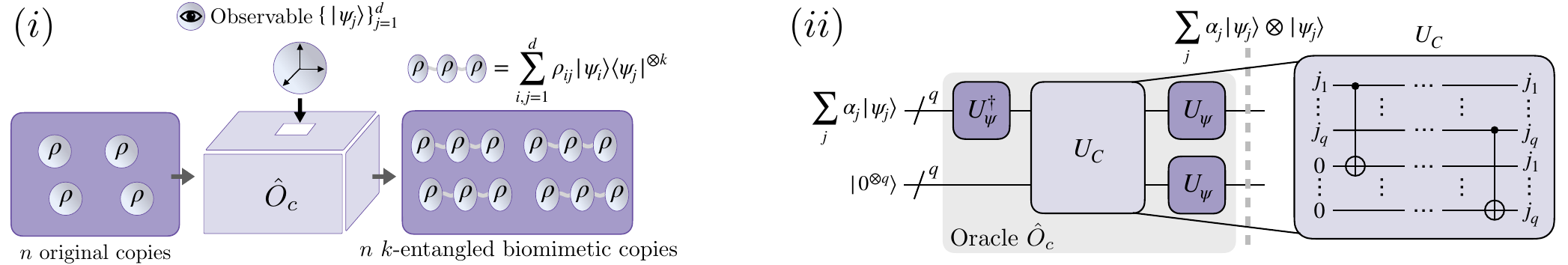}
\caption{$(i)$ Biomimetic cloning machine denoted as the oracle $\hat{O}_c$: requires to chose a preferred basis for the cloning $\{|\psi_j\rangle \}_{j=1}^d$,and performs the operation $|\psi_j\rangle \rightarrow ~|\psi_j\rangle \otimes |\psi_j\rangle$. $(ii)$ Quantum circuit implementation of $\hat{O}_c$. The unitary $U_\psi$ is the change of basis from the computational basis $\{|j\rangle\}_{j=1}^d$ to the preferred basis for the cloning $\{|\psi_j\rangle \}_{j=1}^d$. The structure of the unitary operation $U_C$, which performs the operation $U_C |j\rangle |0\rangle = |j\rangle |j\rangle$ is shown in detail.}
\label{fig:bio_clonning}
\end{figure*}

\subsection{Biomimetic cloning} 
The no-cloning theorem in quantum information \cite{no_clonning} states the impossibility of perfectly copying an arbitrary unknown quantum state. However, cloning a set of orthogonal states is allowed. The biomimetic cloning of quantum observables, referred to as $\hat{O}_c$, leverages this fact by performing the operation 
\begin{equation}\label{U_BCQO}
    \hat{O}_c:\;|\psi_j\rangle|0\rangle\rightarrow |\psi_j\rangle|\psi_j\rangle,
\end{equation}
where  $\{|\psi_j\rangle \}_{j=1}^d$ is the eigenbasis of certain observable, whose statistics will be exactly replicated by the biomimetic copies. Hence, as illustrated in Fig.~\ref{fig:bio_clonning} $(i)$, given a quantum state $\rho = \sum_{i,j=1}^d \rho_{ij} \ket{\psi_i}\langle\psi_j|$, we can generate $k$ biomimetic copies of this state that we denote as
\begin{equation}\label{copies_bio}
    \hat{O}_c^{(k)}(\rho) = \sum_{i,j=1}^d \rho_{ij} \left(\ket{\psi_i}\langle\psi_j|\right)^{\otimes k}\equiv \rho^{(k)}.
\end{equation}
As shown in Fig.~\ref{fig:bio_clonning} $(ii)$, the operation $\hat{O}_c$ can typically be decomposed into three distinct steps. First, we consider the unitary $U_\psi$, which changes from the computational basis $\{|j\rangle\}_{j=1}^d$ to the preferred basis for the cloning $\{|\psi_j\rangle \}_{j=1}^d$ and we perform the transformation
\begin{equation}
    (U_\psi^\dagger\otimes \mathds{1})\;|\psi_j\rangle |0\rangle =|j\rangle |0\rangle.
\end{equation}
Subsequently, we define the unitary $U_{C}$, which replicates the state in the computational basis, i.e.
\begin{equation}
    U_C\;|j_1\dots j_q\rangle|0\dots 0\rangle=|j_1\dots j_q\rangle |j_1\dots j_q\rangle,
\end{equation}
where $j_i \in \{0,1\}$ and $q=\log d$. This corresponds to applying $\mathrm{CNOT}$ gates between the qubits of the two registers, i.e. $U_C=\prod_{i=1}^q U_{\mathrm{CNOT}}^{i,i+q}$, where $U_{\mathrm{CNOT}}^{l,m}$ refers to applying a $\mathrm{CNOT}$ with qubit $l$ acting as control and $m$ as target. In case it is possible to apply the $\mathrm{CNOT}s$ in parallel, the implementation of $U_C$ can be done in a single step, where as if they have to be applied successively and the complexity will be $q$. 
Finally, we undo the change of basis in the two registers
\begin{equation}
    (U_\psi\otimes U_\psi) \; |j\rangle |j\rangle = |\psi_j\rangle |\psi_j\rangle.
\end{equation}
Algorithm \ref{algo_1} provides a comprehensive overview of the full construction of $\hat{O}_c^{(k)}$ for generating $\rho^{(k)}$. Note that an efficient implementation of $U_\psi$ is likely to depend on the chosen preferred basis. Nonetheless, it is worth noting that local or semilocal bases are widespread in practice and do not introduce any hindrance to the implementation of $U_\psi$.

\begin{algorithm}[H]
\caption{Oracle $\hat{O}_c^{(k)}$}\label{algo_1}
\begin{algorithmic}[1]
    \State \textbf{Input:} $\rho$, $k$ ancilla registers in $|0^{\otimes q}\rangle$
    \State \textbf{Assumption:} $U_\psi$ and $U_\psi^\dagger$ can be implemented
    \State Apply $U_\psi^\dagger$ on $\rho$
    \For{$i = 2$ to $k$}
        \State Apply CNOT between the first register and the $i$-th register
    \EndFor
    \State Apply $U_\psi$ on all registers
    \State \textbf{Output:} $\rho^{(k)}$
\end{algorithmic}
\end{algorithm}

\begin{figure*}[t!]
\centering
\includegraphics[width=.92\textwidth]{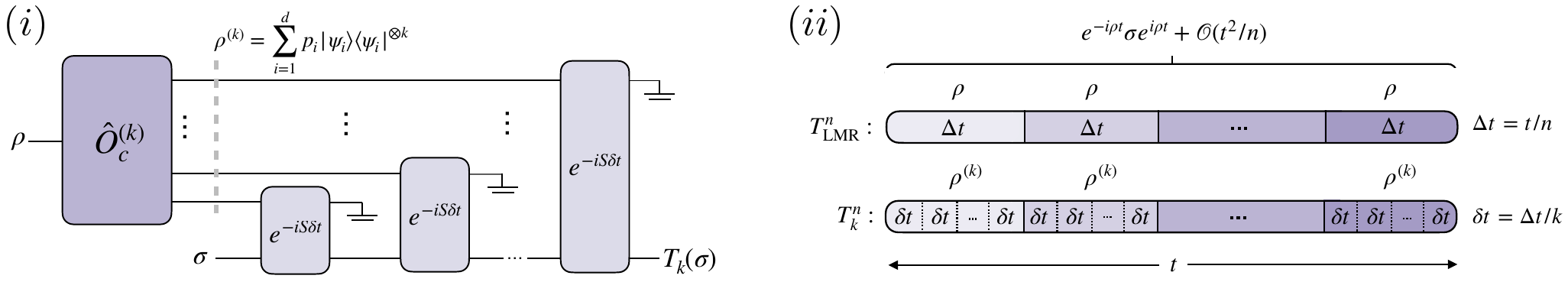}
\caption{$(i)$ LMR protocol for a single original copy of $\rho$ asissted by biomimetic cloning using the eigenbasis of $\rho$ as the preferred basis for the cloning. $S$ denotes the swap operation, $p_i$ the eigenvalues of $\rho$ and $T_k(\sigma)$ the output of the combined operation. $(ii)$ Distribution of time intervals for the combined protocol. To implement the exponentiation of $\rho$ for a time $t$ with $n$ copies, initially, time intervals of duration $\Delta t=t/n$ are considered. When introducing biomimetic copies, each interval is subdivided into $k$ intervals of duration $\delta t=\Delta t/k$.}
\label{fig:circuit}
\end{figure*}

\section{LMR assisted by biomimetic copies} 
We consider a state $\rho$ and a time interval $\Delta t$. We implement the biomimetic cloning on $\rho$ taking an arbitrary observable and generate $k$ biomimetic copies. Thus, the resulting state is the one from Eq.~\ref{copies_bio}.
Next, we apply the LMR protocol, Eq.~\ref{eq:LMR}, to each copy for time intervals of lenght $\delta t= \Delta t/k$ as shown in Fig.~\ref{fig:circuit} $(i)$. This operation is denoted as the quantum channel
\begin{equation}
\label{eq:channel_bio}
    T_k(\sigma) = \tr_{1^{\perp}}\left[\prod_{j=2}^{k+1} e^{-i\delta t S_{1,j}} \sigma \otimes \rho^{(k)} \left(\prod_{j=2}^{k+1} e^{i\delta t S_{1,j}}\right)^{\dag} \right].
\end{equation} 
The resulting state, as fully derived in Appendix~\ref{full_derivation}, is given by
\begin{equation}\label{channel_bio_main}
    \begin{split}
        T_{k}(\sigma) =& \cos^{2k}\delta t\;\sigma - i \cos^k \delta t\;\sin{\Delta t} \;[\rho\circ\mathds{1},\sigma]\\
        &+ (1-\cos^{2k}\delta t) \;\rho\circ\mathds{1}\\
        &+ \cos^k\delta t ( \cos{\Delta t}-\cos^{k}\delta t) \bigl( \{\rho\circ\mathds{1},\sigma\}- 2\; \rho\circ\mathds{1}\circ\sigma \bigr),
    \end{split}
\end{equation}
where $A\circ B=\sum_{i,j}a_{ij}b_{ij}|i\rangle\langle j|$ denotes the Hadamard product. 

\subsection{Preferred basis for the cloning}
To enhance the LMR using biomimetic copies, one should consider a basis that incorporates information relevant to the problem at hand. Examining the desired operation $e^{-i\rho t}\:\sigma\:e^{i\rho t}$, we can identify three bases that have the potential to improve the LMR for matrix exponentiation: the basis that diagonalizes the state $\rho$, the basis that diagonalizes the state $\sigma$, and the basis that diagonalizes some observable $\theta$ whose evolution is of interest. Our study concludes that only considering the eigenbasis of $\rho$ can enhance the performance of the original LMR. To illustrate this, we analyze the regime where a large number $k\gg 1$ of biomimetic cloning operations are performed
\begin{equation}\label{output_channel}
    \begin{split}
        T_k(\sigma) =& \sigma- i \sin{\Delta t} \;[\rho\circ\mathds{1},\sigma]\\
        &+2\sin^2{\Delta t/2} \; \bigl(2\;\rho\circ\mathds{1}\circ\sigma-\{\rho\circ\mathds{1},\sigma\}\bigr)\\&+\frac{\Delta t^2}{2k}\bigl(2(\rho-\sigma)+\{\rho\circ\mathds{1},\sigma\}-2\rho\circ\mathds{1}\circ\sigma\bigr)+\mathcal{O}(\Delta t^3/k).
    \end{split}
\end{equation}
In the first case, where we clone in the eigenbasis of $\rho$, we have $\rho\circ\mathds{1}=\rho$, rendering the protocol resilient against non-diagonal noise in the original copies of $\rho$. Including this in the Eq.~\ref{output_channel} and keeping terms up to second order in $\Delta t$, we obtain
\begin{align*}
    T_k(\sigma) =& \sigma- i \sin{\Delta t} \;[\rho,\;\sigma]+2\sin^2{\Delta t/2} \; \bigl(2\;\rho\circ\sigma-\{\rho,\;\sigma\}\bigr)\\
    &+\mathcal{O}(\Delta t^2/k),
\end{align*}
which is the focus of our work.

In the second case, we use the basis that diagonalizes $\sigma$ for cloning. In this scenario, $[\rho\circ\mathds{1},\sigma]=0$ and $\{\rho\circ\mathds{1},\sigma\}=2(\rho\circ \sigma)$. Therefore, the expression simplifies to
\begin{equation}
    T_{k}(\sigma) = \sigma + \mathcal{O}(\Delta t^2/k).
\end{equation}
Consequently, cloning in this basis converges to the initial state $\sigma$.

In the third case, we use the basis that diagonalizes some observable $\theta$, whose evolution $\tr{\theta e^{-i\rho t}\:\sigma\:e^{i\rho t}}$ is of interest. Cloning in this basis yields the following evolution for the expected value of $\theta$,
\begin{equation}\label{limit_channel}
    \tr(T_{k}(\sigma)\;\theta)= \tr(\sigma\;\theta)+\mathcal{O}(\Delta t^2/k),
\end{equation}
which converges to the expected value with respect to the initial state.

\subsection{Introducing biomimetic copies in the $\rho$ eigenbasis}
Working in the eigenbasis of $\rho$, if we now assume that we have access to $n$ copies of $\rho$, we can generate $k$ biomimetic copies from each original copy and, as depicted in Fig.~\ref{fig:circuit} $(ii)$, we can use them to implement a $\Delta t$ time step of the LMR protocol operation decomposing it into $k$ steps of length $\delta t$. Mathematically, this translates into iteratively applying the operation from Eq.~\ref{output_channel} to each original copy, resulting in the expression 
\begin{equation}\label{full_combined}
\begin{split}
    T_k^{n}(\sigma)=&\sigma-in\Delta t[\rho,\;\sigma]-\frac{n(n-1)}{2}[\rho,\;\sigma]_2\Delta t^2\\
    &-\frac{\Delta t^2}{2}n(\{\rho,\;\sigma\}-2\rho\circ\sigma)\\
    &+\frac{\Delta t^2}{2k}\left(2(\rho-\sigma)+\{\rho,\;\sigma\}-2\rho\circ\sigma\right)+\mathcal{O}(n^2\Delta t^3).
\end{split}
\end{equation}
We assume the ability to generate biomimetic copies, allowing us to operate within the limit $k\xrightarrow{}\infty$ and we define the combined operation as
\begin{equation}
    T_\mathrm{BIO}^n(\sigma)\equiv\lim_{k\rightarrow \infty} T_{k}^{n}(\sigma),
\end{equation}
which is just Eq.~\ref{full_combined}, but removing the $\mathcal{O}(\Delta t^2/k)$ term.

Taking $t=n\Delta t$, the asymptotic error of the LMR protocol assisted with biomimetic copies is given by
\begin{equation}
    \begin{split}
        \text{$\varepsilon$}_{\mathrm{BIO}(n\rightarrow nk)}&=\norm{T_\mathrm{BIO}^n(\sigma) -e^{-i\rho t}\:\sigma\:e^{i\rho t}}_1 \\
    &\approx \frac{t^2}{2n}\norm{\;[\rho,\;\sigma]_2+2\; \rho \circ\sigma-\{\rho,\sigma\}\;}_1\leq \mathcal{O}(t^2/n).
    \end{split}
\end{equation}
A pseudocode outlining the integration of the LMR method with biomimetic copies for performing density matrix exponentiation is provided in Algorithm~\ref{algo_2}.

\begin{algorithm}[H]
\centering
\caption{ $e^{-i \rho t}$ using LMR and the oracle $\hat{O}_c^{(k)}$}\label{algo_2}
\begin{algorithmic}[1]
    \State \textbf{Input:} $\rho^{\otimes n}$, $\sigma$, $t$, $k$, $n$ queries to $\hat{O}_c^{(k)}$
    \State Set $\delta t = t / (nk)$ 
    \For{each copy of $\rho$}
        \State Apply $\hat{O}_c^{(k)}$ to $\rho$ to obtain $\rho^{(k)}$
        \For{$i=1$ to $k$}
            \State Apply $\mathrm{tr}_i\left[e^{-iS_{1i}\delta t}(\sigma\otimes \rho^{(k)})\;e^{iS_{1i}\delta t}\right]$
        \EndFor
    \EndFor
    \State \textbf{Output:} $e^{-i \rho t}\sigma e^{i \rho t}+\mathcal{O}(t^2/n)$
\end{algorithmic}
\end{algorithm}

\begin{figure*}[t!]
\centering
\includegraphics[width=\textwidth]{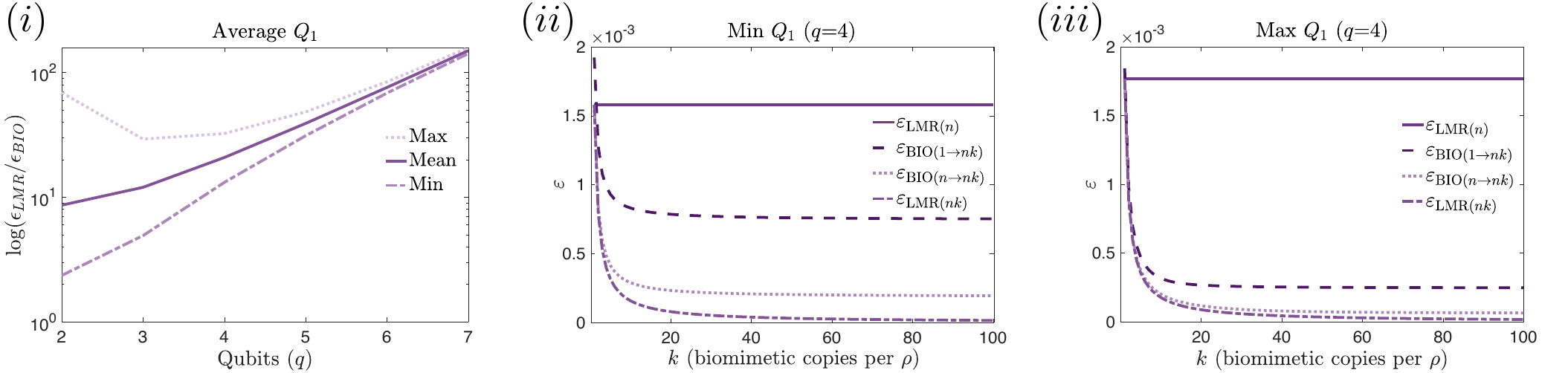}
\caption{We depict the results obtained from randomly generated cases distributed according to the Hilbert-Schmidt measure, with $t=0.2$ and $n=4$ original copies. $(i)$ Mean value of $Q_1$ across a sample of 100,000 random cases for each number of qubits. $(ii)$ Minimum $(ii)$ and maximum $(iii)$ value of $Q_1$ among the 100,000 random samples, specifically for $q=4$.}
\label{fig:results}
\end{figure*}

\subsection{Performance evaluation and numerical results}
The convergence behavior of the error with respect to the number of biomimetic copies, denoted as $k$, is shown in Figs.~\ref{fig:results} $(ii)$ and $(iii)$, where we show the best and worst cases respectively. These graphics present the error in different scenarios. Firstly, the error is depicted when directly applying LMR with the original $n$ copies, which corresponds to Eq.~\ref{error_LMR}. Additionally, the graphics show the error for two cases of the combined protocol: one starting with 1 original copy and generating $nk$ biomimetic copies, and the other starting with $n$ original copies and generating $nk$ biomimetic copies. Finally, the error is also displayed for the LMR with $nk$ original copies. When provided with $n$ copies of $\rho$, the improvement achieved by generating biomimetic copies versus the direct application of the LMR operation, Eq. (\ref{error_LMR}), is quantified by 
\begin{equation}\label{Q1}
    \frac{\text{$\varepsilon$}_{\mathrm{LMR}(n)}}{\text{$\varepsilon$}_{\mathrm{BIO}(n\rightarrow nk)}} \approx \frac{\norm{\;[\rho,\;\sigma]_2+2(\rho-\sigma)\;}_1}{\norm{\;[\rho,\;\sigma]_2+2\; \rho \circ\sigma-\{\rho,\sigma\}\;}_1}\equiv Q_1.
\end{equation}
Here, the subscript $1$ in $Q$ denotes the norm-1 or trace norm. Therefore, although the error is suppressed in the same way with the number of copies $n$, thus not violating the optimality result of the original protocol from Ref.~\cite{Kimmel_2017}, the prefactors depend on the quantum states under consideration and can lead to a reduction in the error as the system size increases. Looking at Figs.~\ref{fig:results} $(ii)$ and $(iii)$, $Q_1$ corresponds to the ratio between the horizontal $\text{$\varepsilon$}_{\mathrm{LMR}(n)}$ line and the $\text{$\varepsilon$}_{\mathrm{BIO}(n\rightarrow nk)}$ line once it has converged. Moreover, in Appendix~\ref{demo_2}, we demonstrate that using the Frobenius norm to quantify the error allows us to establish the lower bound $Q_2\geq 2$, with $Q_2$ defined accordingly to Eq.~\ref{Q1} but using the Frobenius norm. The minimum value of this lower bound is actually reached for an specific single qubit configuration, which corresponds to the maximally mixed state for $\rho$. It is worth noting that the denominator exhibits at least a second-order dependence on $\rho$ and $\sigma$, while the numerator only possesses first-order dependence. As a result, when dealing with full rank systems, $Q_1$ will generally have a large value. This behaviour is experimentally depicted in Fig.~\ref{fig:results} $(i)$, where the mean, minimum and max value of $Q_1$ from a sample of 100,000 random cases uniformly distributed according to the Hilbert-Schmidt measure is illustrated for different number of qubits. These numerical results show a linear scaling of $Q_1$ with the dimension of the system for the average case.

\section{Generating vs. cloning $\rho$}
Our protocol offers a significant advantage based on the underlying assumption that either the process for producing the state is unknown, or the generating step incurs a substantially higher cost compared to obtaining biomimetic copies of $\rho$. Here we explore the validity of these assumptions and the applicability of our model.

Let us consider two distinct scenarios. In the first scenario, the complete description of the state $\rho$ is unknown (only its eigenbasis is given), while in the second, the full description of $\rho$ is available.

\subsection{Full description of $\rho$ is unknown}
In this case, we can assume that $\rho$ is obtained as an output of an experiment or from a costly computation. We described this process by the oracle $\hat{O}_s$, which in an optimistic situation can be reproduced. 
In this section, we delineate the regime in which it is advantageous to produce biomimetic copies, taking into consideration both the implementation cost of the biomimetic cloning machine and the cost associated with generating an original copy of $\rho$.

To do this, we start by using Eq.~\ref{output_channel} to get the error of the combined protocol without taking the limit $k\to\infty$,  obtaining the following upper bound
\begin{equation}
\begin{split}
    \text{$\varepsilon$}_{\mathrm{BIO}(n\rightarrow nk)}=&\norm{T_\mathrm{BIO}^n(\sigma) -e^{-i\rho t}\:\sigma\:e^{i\rho t}}_1 =\\
    \leq& \norm{B}_1\frac{n\Delta t^2}{2}+\norm{C}_1\frac{\Delta t^2}{2k}=\norm{B}_1\frac{t^2}{2n}+\norm{C}_1\frac{t^2}{2kn^2},
\end{split}
\end{equation}
where we defined $B\equiv[\rho,\;\sigma]_2+2\; \rho \circ\sigma-\{\rho,\sigma\}$ and $C\equiv 2(\rho-\sigma)+\{\rho,\sigma\}-2\; \rho \circ\sigma$. Now, we consider generating $l$ additional original copies, and applying the original LMR has an associated error
\begin{equation}
    \text{$\varepsilon$}_{\mathrm{LMR}(n+l)}=\norm{A}_1\frac{t^2}{2(n+l)}+\mathcal{O}\left(t^3/(n+l)^2\right),
\end{equation}
where $A\equiv [\rho,\;\sigma]_2+2(\rho-\sigma)=B+C$. In this way, we can calculate the minimum number of additional copies $l$ that we need to achieve an error smaller than or equal to the one obtained if we generate $k$ biomimetic copies from each of the $n$ original copies, obtaining
\begin{equation}
    l\geq n\left( \frac{nk\norm{A}_1}{nk\norm{B}_1+\norm{C}_1} -1\right).
\end{equation}
This lower bound tends to a constant value when considering large $k$. This observation is intuitive: by increasing the number of original copies $l$ to infinity, you can reduce the error to 0. Conversely, with the number of biomimetic copies, you converge to a constant value.

If we denote the cost of generating an extra original copy as $C_{c}$ and the cost of generating a biomimetic copy as $C_{s}$, we aim to determine the regime in which $l\cdot C_s\geq n\cdot (k-1)\cdot C_c$ holds, while using the previous condition for $l$. This yields the condition
\begin{equation}
    C_{c}\leq \frac{C_{s}}{k-1} \frac{nk\norm{A}_1-nk\norm{B}_1-\norm{C}_1}{nk\norm{B}_1+\norm{C}_1}.
\end{equation}
Therefore, whenever the ratio $C_c/C_s$ between the cost of approximated cloning and the cost of the source is small enough such that the latter bound holds, our proposal incurs less cost for the same performance. Now we can analyze the behavior of this last expression in the $nk$ large regime, obtaining the upper bound 
\begin{equation}
    C_{c}\leq\frac{C_{s}}{k-1}\left(\frac{\norm{A}_1}{\norm{B}_1}-1\right)=\frac{C_{s}}{k-1}\left(Q_1-1\right),
\end{equation}
where $Q_1$ as defined in Eq.~\ref{Q1} is the ratio between the error of the LMR protocol with a fixed number of original copies and the error of the protocol assisted with asymptotically many biomimetic copies. The bound shows that when the enhancement in accuracy described by $Q_1$ is large, there is a wider regime for $C_c/C_s$ such that the cost of assisting the protocol with biomimetic copies is smaller than generating as many source copies as to match the performance. The latter bound is approximately tight when the total number of biomimetic copies $nk$ is large, meaning that its fulfillment guarantees less cost by assisting with biomimetic copies. In general, this bound being violated would imply that making new source copies is more efficient than assisting with biomimetic copies.

\subsection{Full description of $\rho$ is known}
In the second scenario, when the description of $\rho$ is fully known, the unitary operation $e^{-i\rho t}$ can be achieved straightforwardly. 
Up to an efficient change of basis, we can assume that we are working in $\rho'$s eigenbasis so $\rho=\sum_j p_j |j\rangle\langle j|$ and $\sigma=|\psi\rangle\langle \psi|$ is a pure state with
$|\psi\rangle=\sum_j \psi_j |j\rangle.$ With this notation our target operation can be written as

\begin{equation}\label{desired}
   e^{-i\rho t} \sigma:  \sum_j \psi_j|j\rangle\xrightarrow{}\sum_j e^{-i p_j t}\psi_j|j\rangle.
\end{equation}

We can now distinguish between two cases: either the eigenvalues $p_j$ are given by a known analytical expression beforehand as a function of the position they occupy $p(j)$, or they are not. In the latter case, we can always construct an interpolation polynomial on the $2^q$ values to achieve an analytical expression. Notice that the classical cost of obtaining the polynomial interpolation scales as $\mathcal{O}(2^{2q})$ and potentially introduce a significant cost overhead.



Let us now analyze how to reach the transformation $e^{-i\rho t}$ given $p(j)$ by using oracles and ancillary qubits, although this is not the only method existing in the literature \cite{Welch_2014}. Firstly, we will assume that we have access to an oracle $\hat{O}_1$ which performs the following transformation
\begin{equation}
    \sum_j \psi_j |j\rangle \otimes |0^{\otimes m}\rangle \xrightarrow{\hat{O}_1} \sum_j \psi_j
    |j\rangle \otimes |p(j)\rangle,
\end{equation}
where $|p(j)\rangle$ denotes the binary codification of $p(j)$ with $m-$bit precision on $m$ additional ancillary qubits up to a certain error $\epsilon$. This leads to the  complexity of calculating $p(j)$ with precision $\epsilon$ growing as $\log(1/\epsilon)$.
The implementation of this oracle might be obtained by translating a classic circuit that calculates $p(j)$, and implement it in reversible form using the standard coherent arithmetic techniques \cite{takahashi2009quantum}. Now, we consider a second oracle $\hat{O}_2$ which introduces the corresponding phase 
\begin{equation}
    |j\rangle \otimes |p(j)\rangle \xrightarrow{\hat{O}_2} e^{-ip(j)t}|j\rangle \otimes |p(j)\rangle.
\end{equation}
This oracle can be attained by adding an extra ancillary qubits and performing $m$ Rz-controlled rotations of angle $1/2^k$, $k=1,\ ... m$ depending on the value of the ancillas encoding the $m-$ bit precision value of $p(j)$ on the ancilla qubit. Next, we post select the value of the ancilla in the state $\ket{0}$ with probability $50\%$. Finally, to obtain the desired operation from Eq.~(\ref{desired}), all we have to do is to apply $\hat{O}_1^\dagger$ and discard the ancillary qubits.

\section{Illustrative examples}
We have shown that the performance of the combined protocol in a general case is characterised by $Q_1$. Let us now consider two specific cases which are of particularly interest to highlight when this method is particularly favorable.

Let us start with the simplest scenario, the single qubit case. As demonstrated in Appendix~\ref{demo_2}, we have shown that the minimum value for both $Q_1$ and $Q_2$ is 2. This implies a reduction by half in the number of original copies required to achieve the desired accuracy. The specific case that yields these minimum values corresponds to $\rho=\mathds{1}/2$ and $\sigma=\mathds{1}/2+c\hat{\sigma}_x+d\hat{\sigma}_y$, where $\hat{\sigma}_x$ and $\hat{\sigma}_y$ are the Pauli matrices and $c$ and $d$ are real numbers chosen such that $\sigma$ is a positive semi-definite matrix. The proof can be found in Appendix~\ref{examples}.

We also consider the scenario when $\rho$ is highly mixed, i.e.
\begin{equation}
    \rho=\sum_{i=1}^d \varepsilon_i |i\rangle\langle i|,
\end{equation}
with $d=2^q$ and $0\leq \varepsilon_i \leq 4/d$, providing a connection to the statistical case, as explained in the following paragraph. In this case, we can obtain the following inequality
\begin{equation}\label{scaling_particular_main}
    Q_1\geq \frac{d}{8}\left(\frac{\norm{\rho-\sigma}_1-32/d^2}{1+4/d}\right),
\end{equation}
with the full derivation provided in Appendix~\ref{examples}. Thus, if $\norm{\rho-\sigma}_1\gg 1/d$, then $Q_1\propto d$. This happens, for instance, when considering $\mathrm{rank}(\sigma)=R<d$. In that case it can be shown that
\begin{equation}
    \norm{\rho-\sigma}_1\geq 1- R\frac{4}{d}.
\end{equation}
and therefore, the exponential scaling of $Q_1$ occurs when $R\ll~d/4$.

We would like to remark that this specific example bears a connection with the statistical average case, providing an insight into the exponential scaling observed in Fig.~\ref{fig:results} $(iii)$. In their work \cite{Pucha_a_2016}, Pucha\l{}a et al. utilized the fact that the eigenvalues of a random density matrix distributed according to the Hilbert-Schmidt measure, converge in probability to the Marchenko-Pastur distribution, which is supported between 0 and $4/d$ (which in fact aligns with the scenario in our example). Through this observation, they demonstrated that the trace distance between two random density matrices converges to a constant value. Consequently, when we apply this finding to Eq.~(\ref{scaling_particular_main}), we can infer that $Q_1$ exhibits a linear scaling with $d$ in the average case.


\section{Conclusions} 
In this article we have proposed an enhanced density matrix exponentiation protocol by combining the LMR with the introduction of imperfect copies of the quantum state by means of the biomimetic cloning machine. This approach significantly improves the protocol's performance when limited copies are available, assuming that the eigenbasis of $\rho$ is known. Given an error $\epsilon$, our methodology reduces drastically the required number of original copies with respect to the original LMR for larger system sizes. To support our proposal, we have presented both analytical and numerical arguments demonstrating that, despite the original protocol being optimal in the number of copies $n$, our method can show an average enhancement that scales exponentially with the size of the system. We have also discussed the scenarios in which employing our protocol enables to achieve an speed up, comparing to either generating the quantum state $\rho$ directly or directly performing its exponentiation.

As a final remark, even though we have focused our study in a particular non-linear transformation, the exponentiation, our results constitute a promising starting point to enhance the implementation of many quantum protocols that require multiple copies of a quantum state, particularly when its breed resources are costly or limited.

\section*{acknowledgements}
The Authors acknowledge support from EU FET Open project EPIQUS (899368) and HORIZON-CL4- 2022-QUANTUM01-SGA project 101113946 OpenSuperQPlus100 of the EU Flagship on Quantum Technologies, the Spanish Ramón y Cajal Grant RYC-2020-030503-I, project Grant No. PID2021-125823NA-I00 funded by MCIN/AEI/10.13039/501100011033 and by “ERDF A way of making Europe” and “ERDF Invest in your Future”, and from the IKUR Strategy under the collaboration agreement between Ikerbasque Foundation and BCAM on behalf of the Department of Education of the Basque Government.This project has also received support from the Spanish Ministry for Digital Transformation and of Civil Service of the Spanish Government through the QUANTUM ENIA project call - Quantum Spain, EU through the Recovery, Transformation and Resilience Plan – NextGenerationEU within the framework of the Digital Spain 2026. We acknowledge funding from Basque Government through Grant No. IT1470-22. P.R-G acknowledges the CDTI within the Misiones 2021 program and the Ministry of Science and Innovation under the Recovery,
Transformation and Resilience Plan—Next Generation EU under the project “CUCO: Quantum Computing
and its Application to Strategic Industries”. R.I. acknowledges the support of the Basque Government Ph.D. grant PRE 2021-1-0102. J.G-C acknowledges the support from the UPV/EHU Ph.D. Grant No. PIF20/276. PR is supported by the National Research Foundation, Singapore, and A*STAR under its CQT Bridging Grant. YB acknowledges support from the Spanish Government via the project PID2021-126694NA- C22 (MCIU/AEI/FEDER, EU) and the Ramon y Cajal grant (RYC2023-042699-I).

\begin{appendix}
\section{Mathematical formulation of the combined protocol}\label{full_derivation}
The objective of this section is to explore the mathematical formulation of the channel of the combined protocol and to examine various potential choices for the preferred basis for the biomimetic cloning. The operation that describes biomimetic cloning along with the application of the LMR is given by the following expression
\begin{equation}
    T_k(\sigma) = \tr_{1^{\perp}}\left[\prod_{j=2}^{k+1} e^{-i\delta t S_{1,j}} \sigma \otimes \rho^{(k)} \left(\prod_{j=2}^{k+1} e^{i\delta t S_{1,j}}\right)^{\dag} \right].
\end{equation}
Here, $\rho^{(k)} = \sum_{i,j=1}^d \rho_{ij} |\psi_i\rangle \langle \psi_j|^{\otimes k}$ represents the biomimetic copies, which can be entangled when considering a general cloning basis. When $k=1$, the case corresponds to $\rho^{(1)}=\rho$, and we retrieve the LMR with a single copy. Let us define $\abs{x}=\sum_{j=1}^k x_j$ as the number of ones in the bitstring $x$ and express the product of partial swaps with trigonometric functions
\begin{equation}
\begin{split}
    \prod_{j=2}^{k+1} e^{-i\delta t S_{1,j}} &=
    \prod_{j=2}^{k+1} (\cos\delta t \mathds{1} - i \sin \delta t S_{1, j}) \\
    &=
    \sum_{x\in\qty{0,1}^k} (\cos \delta t)^{k-\abs{x}} (-i \sin \delta t)^{\abs{x}} \prod_{j=1}^{k} S_{1,j+1}^{x_j}.
\end{split}
\end{equation}
We obtain
\begin{equation}
\begin{split}
    T_{k}(\sigma) = &
    \sum_{x\in\qty{0,1}^k}\sum_{y\in\qty{0,1}^k} (\cos \delta t)^{2k-\abs{x}-\abs{y}} (-i \sin \delta t)^{\abs{x}} (i \sin \delta t)^{\abs{y}}\\
    &\cdot\tr_{1^{\perp}}\qty[\prod_{j=1}^{k} S_{1,j+1}^{x_j} \sigma \otimes \rho^{(k)} \qty(\prod_{i=1}^{k} S_{1,i+1}^{y_i})^{\dag} ].
\end{split}
\end{equation}
This expression can be simplified by introducing the representation of $\sigma$ in the preferred basis for the cloning $\sigma_{ij}= \langle\psi_i|\sigma|\psi_j\rangle$ and by expressing as $\rho\circ\sigma = \sum_{i,j=1}^d \rho_{ij}\sigma_{ij}|\psi_i\rangle\langle \psi_j|$, i.e. the Hadamard product (element-wise product) between the corresponding density matrices represented in the preferred basis for the cloning:
\begin{enumerate}[label={(\arabic*)}]
    \item $\tr_{1^{\perp}}[S_{1j_m}... S_{1j_1} \sigma \otimes \rho^{(k)} S_{1i_1}...S_{1i_l}] = \tr_{1^{\perp}}[S_{1j_1} \sigma \otimes \rho^{(k)} S_{1,i_1}]$, if the sequences $j_1,...,j_m$ and $i_1,...,i_l$ have no repetitions.
    \item $\tr_{1^{\perp}}[S_{1j} \sigma \otimes \rho^{(k)}] = \sum_{i,l=1}^d \rho_{ii}\sigma_{li}|\psi_l\rangle\langle \psi_i|=\sigma\;(\rho\circ\mathds{1})$ and $\tr_{1^{\perp}}[\sigma \otimes \rho^{(k)} S_{1j}] = \sum_{i,l=1}^d \rho_{ii}\sigma_{il}|\psi_l\rangle\langle \psi_i|=(\rho\circ\mathds{1})\;\sigma$ for $j= 2,...,k$. 
    \item $\tr_{1^{\perp}}[S_{1j_m} ... S_{1j_1} \sigma \otimes \rho^{(k)}] = \sigma\;(\rho\circ\mathds{1})$ and $\tr_{1^{\perp}}[\sigma \otimes \rho^{(k)} S_{1j_m} ... S_{1j_1}] = (\rho\circ\mathds{1})\;\sigma$ if the sequence $j_1,...,j_m$ has no repetitions.
    \item $\tr_{1^{\perp}}[S_{1j} \sigma \otimes \rho^{(k)} S_{1j}] = \tr{\sigma} \sum_{i=1}^d\rho_{ii}|\psi_i\rangle\langle \psi_i|=\tr{\sigma}\;(\rho\circ\mathds{1})$. 
    \item $\tr_{1^{\perp}}[S_{1j} \sigma \otimes \rho^{(k)} S_{1i}] = \sum_{p=1}^d \rho_{pp}\sigma_{pp}|\psi_p\rangle\langle \psi_p|=\rho\circ\mathds{1}\circ\sigma$. 
\end{enumerate}
Introducing the first relation and defining $\gamma(x) := \min \qty{j\in \{1,2,\ldots,k\}  \;|\; x_j = 1}$ to denote the index of the first 1 in the non-zero bit string $x\neq 0:= \{0,\ldots,0\}$, the operator $T_k$ reads
\begin{align*}
    T_{k}(\sigma) &= (\cos \delta t)^{2k} \sigma \\
    & + \sum_{y\in\qty{0,1}^k\setminus \qty{0}} (\cos \delta t)^{2k-\abs{y}} (i \sin \delta t)^{\abs{y}} \tr_{1^{\perp}}[\sigma \otimes \rho^{(k)} S_{1,\gamma(y)+1} ] \\
    & + \sum_{x\in\qty{0,1}^k\setminus \qty{0}} (\cos \delta t)^{2k-\abs{x}} (-i \sin \delta t)^{\abs{x}} \tr_{1^{\perp}}[S_{1,\gamma(x)+1} \sigma \otimes \rho^{(k)}] \\
    & + \sum_{x\in\qty{0,1}^k\setminus \qty{0}}\sum_{y\in\qty{0,1}^k\setminus\qty{0}} (\cos \delta t)^{2k-\abs{x}-\abs{y}} (-i \sin \delta t)^{\abs{x}} (i \sin \delta t)^{\abs{y}} \\
    &\phantom{\sum_{x\in\qty{0,1}^k\setminus \qty{0}}\sum_{y\in\qty{0,1}^k\setminus\qty{0}}}\cdot\tr_{1^{\perp}}[S_{1,\gamma(x)+1} \sigma \otimes \rho^{(k)} S_{1,\gamma(y)+1} ].
\end{align*}
Employing the remaining relations
\begin{equation}\label{pre_sums_channel}
\begin{split}
    T_{k}(\sigma) &= (\cos \delta t)^{2k} \sigma \\
    & + \qty(\sum_{y\in\qty{0,1}^k\setminus \qty{0}} (\cos \delta t)^{2k-\abs{y}} (i \sin \delta t)^{\abs{y}})\; \sigma\; (\rho\circ\mathds{1}) \\
    & + \qty(\sum_{x\in\qty{0,1}^k\setminus \qty{0}} (\cos \delta t)^{2k-\abs{x}} (-i \sin \delta t)^{\abs{x}})  \;(\rho\circ\mathds{1})\;\sigma \\
    & + \sum_{x\in\qty{0,1}^k\setminus \qty{0}}\sum_{y\in\qty{0,1}^k\setminus\qty{0}} (\cos \delta t)^{2k-\abs{x}-\abs{y}} (-i \sin \delta t)^{\abs{x}} (i \sin \delta t)^{\abs{y}} \\
    & \phantom{aaaaa}\cdot\qty[ \delta_{\gamma(x),\gamma(y)} \tr{\sigma}\;(\rho\circ\mathds{1}) (1-\delta_{\gamma(x),\gamma(y)}) (\rho\circ\mathds{1}\circ\sigma)].
\end{split}
\end{equation}
We can compute the summations in the first and second terms noting that there are $\mqty(k\\m)$ bitstrings of size $k$ with $\abs{x}=m$ and the binomial expansion $(a+b)^k=\sum_{m=0}^k \mqty(k\\m)a^{k-m}b^m$
\begin{align*}
\sum_{y\in\qty{0,1}^k\setminus\qty{0}} (\cos \delta t)^{2k-\abs{y}} (i \sin \delta t)^{\abs{y}} &= \sum_{m=1}^k \mqty(k\\ m) (\cos\delta t)^{2k-m} (i \sin \delta t)^m \\
&= (\cos\delta t)^k \qty(e^{i k \delta t} - (\cos\delta t)^k)\\
\sum_{x\in\qty{0,1}^k\setminus\qty{0}} (\cos \delta t)^{2k-\abs{x}} (-i \sin \delta t)^{\abs{x}} &= \sum_{m=1}^k \mqty(k\\ m) (\cos\delta t)^{2k-m} (-i\sin \delta t)^m \\
&= (\cos\delta t)^k \qty(e^{-i k \delta t} - (\cos\delta t)^k)
\end{align*}
For the summation in the third term, we first consider the case where $\gamma(x)=\gamma(y)$. We note that there are $\mqty(k-\gamma\\ m-1)$ possible bitstrings $x$ satisfying $\gamma (x) = \gamma \in\qty{1,...,k}$ and $\abs{x}=m\in\qty{1, ..., k-\gamma+1}$ and proceed as follows
\begin{align*}
&\sum_{x\in\qty{0,1}^k\setminus \qty{0}}\sum_{y\in\qty{0,1}^k\setminus\qty{0}} (\cos \delta t)^{2k-\abs{x}-\abs{y}} (-i \sin \delta t)^{\abs{x}} (i \sin \delta t)^{\abs{y}} \delta_{\gamma(x),\gamma(y)}\\ 
&= \sum_{\gamma=1}^k \sum_{m=1}^{k-\gamma+1} \sum_{l=1}^{k-\gamma+1} \mqty(k-\gamma\\ m-1)\mqty(k-\gamma\\l-1) (\cos \delta t)^{2k-m-l} (-i \sin \delta t)^{m} (i \sin \delta t)^{l} \\
&= \sum_{\gamma=1}^n \qty[\sum_{m=1}^{k-\gamma+1} \mqty(k-\gamma\\ m-1) (\cos \delta t)^{k-m} (-i \sin \delta t)^{m}]\\
&\phantom{= \sum_{\gamma=1}^n}\cdot\qty[ \sum_{l=1}^{k-\gamma+1} \mqty(k-\gamma\\l-1) (\cos \delta t)^{k-l}(i \sin \delta t)^{l}],
\end{align*}
computing
\begin{align*}
    &\sum_{m=1}^{k-\gamma+1} \mqty(k-\gamma\\ m-1) (\cos \delta t)^{k-m} (-i \sin \delta t)^{m} \\
    &= (\cos \delta t)^{\gamma-1} (-i\sin\delta t)\sum_{m=1}^{k-\gamma+1} \mqty(k-\gamma\\ m-1) (\cos \delta t)^{k-\gamma-(m-1)} (-i \sin \delta t)^{m-1} \\
    &= (\cos \delta t)^{\gamma-1} (-i\sin\delta t) e^{-i(k-\gamma) \delta t},
\end{align*}
we obtain
\begin{align*}
&\sum_{x\in\qty{0,1}^k\setminus \qty{0}}\sum_{y\in\qty{0,1}^n\setminus\qty{0}} (\cos \delta t)^{2n-\abs{x}-\abs{y}} (-i \sin \delta t)^{\abs{x}} (i \sin \delta t)^{\abs{y}} \delta_{\gamma(x),\gamma(y)}\\ 
&= \sum_{\gamma=1}^k (\cos \delta t)^{2\gamma-2} (\sin\delta t)^2 e^{-i(k-\gamma) \delta t} e^{i(k-\gamma) \delta t} \\
&=  (\sin\delta t)^2 \sum_{\gamma=1}^k (\cos \delta t)^{2(\gamma-1)} \\
&= (\sin\delta t)^2 \frac{1-(\cos \delta t)^{2k}}{1-(\cos \delta t)^2}\\
&= 1-\cos^{2k} \delta t,
\end{align*}
where we have introduced the closed-form formula for the partial sum of the geometric series $\sum_{r=0}^{k-1} a^r = \frac{1-a^k}{1-a}$. Finally, we consider the case where $\gamma(x)\neq\gamma(y)$ and obtain
\begin{align*}
    &\sum_{x,y\in\qty{0,1}^k\setminus \qty{0}} (1-\delta_{\gamma(x),\gamma(y)})(\cos \delta t)^{2k-\abs{x}-\abs{y}} (-i \sin \delta t)^{\abs{x}} (i \sin \delta t)^{\abs{y}} \\
    &= \sum_{\alpha=1}^k \sum_{\beta=1}^k (1-\delta_{\alpha,\beta}) 
    (\cos \delta t)^{\alpha-1} (-i\sin\delta t) e^{-i(k-\alpha) \delta t}\\
    &\phantom{= \sum_{\alpha=1}^k \sum_{\beta=1}^k}\cdot(\cos \delta t)^{\beta-1} (i\sin\delta t) e^{i(k-\beta) \delta t}\\
    &= \sin^2 \delta t \sum_{\alpha=1}^k \sum_{\beta=1}^k (1-\delta_{\alpha,\beta}) 
    (\cos \delta t)^{\beta+\alpha-2} e^{-i(\beta-\alpha) \delta t}\\
    &= 2\cos^{k} \delta t\;(\cos^k \delta t - \cos k\delta t),
\end{align*}
Replacing these sums, Eq.~\ref{pre_sums_channel} reduces to
\begin{equation}\label{channel_bio_apendix}
\begin{split}
    T_{k}(\sigma) = &(\cos\delta t)^{2k} \;\sigma - i (\cos \delta t)^k \;\sin{k\delta t}\; [\rho\circ\mathds{1},\sigma]\\
    &+ (1-\cos^{2k}\delta t) \;\tr{\sigma}\;(\rho\circ\mathds{1})\\
    &+ (\cos\delta t)^k (\cos{k\delta t}-\cos^k\delta t) \qty(\{\rho\circ\mathds{1},\sigma\}-2\;\rho\circ\mathds{1}\circ\sigma).
\end{split}
\end{equation}

The expression above describes the action of the LMR protocol with one original copy $\rho$ assisted by $k$ applications of the biomimetic cloning in an arbitrary basis $\qty{|\psi_i\rangle}_{i=1}^d$.

\section{Lower bound of the error quotient}\label{demo_2}
In this section we will prove a lower bound for the quotient of the errors. To simplify the analytical calculations, we will consider the squared Frobenius norm $\norm{A}_2^2=\mathrm{tr}(AA^\dagger)$ instead of the trace norm. This way, we define the quotient
\begin{equation}\label{inequality}
    Q_2^2\equiv\frac{\norm{\;[\rho,\;\sigma]_2+2(\rho-\sigma)\;}_2^2}{\norm{\;[\rho,\;\sigma]_2+2\; \rho \circ\sigma-\{\rho,\sigma\}\;}_2^2}\geq 4.
\end{equation}
To prove this lower bound, as $\rho$ is diagonal, it is useful to separate $\sigma=D+X$, where $D$ is the matrix which contains the diagonal elements of $\sigma$ and $X$ the non-diagonal ones. This way we can take advantage of the fact that $\rho$ does commute with $D$ and simplify
\begin{equation}
    \frac{\norm{\;[\rho,\;\sigma]_2+2(\rho-\sigma)\;}_2^2}{\norm{\;[\rho,\;\sigma]_2+2\; \rho \circ\sigma-\{\rho,\sigma\}\;}_2^2}=\frac{\norm{\;[\rho,\;X]_2+2(\rho-D-X)\;}_2^2}{\norm{\;[\rho,\;X]_2-\{\rho,X\}\;}_2^2}.
\end{equation}
We can utilize the following inequality
\begin{equation}\label{first_inequality}
    \frac{\norm{\;[\rho,\;X]_2+2(\rho-D-X)\;}_2^2}{\norm{\;[\rho,\;X]_2-\{\rho,X\}\;}_2^2}\geq \frac{\norm{\;[\rho,\;X]_2-2X\;}_2^2}{\norm{\;[\rho,\;X]_2-\{\rho,X\}\;}_2^2},
\end{equation}
which is sharp. To demonstrate its validity, we introduce the definitions $X'\equiv [\rho,\;X]_2-2X$ which is a non-diagonal matrix, and $D'\equiv 2(\rho-D)$ which is a diagonal real matrix. Thus, the inequality can be expressed as
\begin{equation}
    \norm{X'+D'}_2^2\geq \norm{X'}_2^2.
\end{equation}
which, employing the Frobenius norm definition, is equivalent to $\mathrm{tr}(D'^2)+2\mathrm{tr}(X'D')\geq 0$. Here, $\mathrm{tr}(X'D')=0$. Thus, we are left with $\mathrm{tr}(D'^2)\geq 0$, which clearly holds true.
\\
\\
Now we compute the Frobenius norm squared of numerator
\begin{align*}
&\norm{\;[\rho,\;X]_2-2X\;}_2^2\\
&\phantom{aaaa}=\mathrm{tr}\Bigl(\sum_{i,j=1}^d \bigl(2-(p_i-p_j)^2\bigr)\;x_{ij}\;\sum_{k,l=1}^d \bigl(2-(p_k-p_l)^2\bigr)\;x_{kl}\Bigr) \\
&\phantom{aaaa}=\sum_{k,r=1}^d\Bigl(2-(p_k-p_r)^2\Bigr)^2|x_{kr}|^2,
\end{align*}
and denominator
\begin{align*}
    &\norm{\;[\rho,\;X]_2-\{\rho,X\}\;}_2^2=\mathrm{tr}\Bigl(\sum_{i,j=1}^d \bigl(p_i+p_j-(p_i-p_j)^2\bigr)\;x_{ij}\\
    &\phantom{\norm{\;[\rho,\;X]_2-\{\rho,X\}\;}_2^2=\mathrm{tr}\Bigl(}\cdot\sum_{k,l=1}^d \bigl(p_k+p_l-(p_k-p_l)^2\bigr)\;x_{kl}\Bigr) \\
    &\phantom{aaaa} =\sum_{k,r=1}^d\Bigl(p_k+p_r-(p_k-p_r)^2\Bigr)^2|x_{kr}|^2,
\end{align*}
where as $X$ is Hermitian we used $x_{kr}x_{rk}=x_{kr}\Bar{x}_{kr}=|x_{kr}|^2$. The inequality is now explicitly expressed as follows
\begin{equation}\label{quotient}
    Q_2^2\geq \frac{\sum_{k,r=1}^d\Bigl(2-(p_k-p_r)^2\Bigr)^2|x_{kr}|^2}{\sum_{k,r=1}^d\Bigl(p_k+p_r-(p_k-p_r)^2\Bigr)^2|x_{kr}|^2}\geq 4,
\end{equation}
or equivalently
\begin{equation}\label{simplified_ineq}
\begin{split}
    \sum_{k,r}\Bigl(&4+(p_k-p_r)^4-4(p_k-p_r)^2-4(p_k+p_r)^2\\
    &+4(p_k-p_r)^4+2(p_k+p_r)(p_k-p_r)^2\Bigr)|x_{kr}|^2\geq 0.
    \end{split}
\end{equation}
This can be written more compactly as $\sum_{kr}M_{kr}|x_{kr}|^2\geq 0$, where $M_{kr}$ represents the combined coefficients. To prove this inequality, we will show that the coefficients $M_{kr}\geq 0$ for all $k$ and $r$. We define the variables 
\begin{align}
    &p_+=p_k+p_r,\\
    &p_-=(p_k-p_r)^2,
\end{align}
and, since $x_{kk}=0$, we do not consider having $k=r$. This implies that $p_k+p_r\leq 1$, ensuring that the conditions 
\begin{equation}\label{conditions}
    0\leq p_-\leq p_+\leq 1
\end{equation}
are satisfied. By substituting these definitions and rearranging terms, we obtain
\begin{equation}
    M_{kr}=4(1-p_+^2)+(8p_+-3p_--4)p_-.
\end{equation}
We can check that, using the conditions from Eq.~\ref{conditions}, 
\begin{equation}\label{minimum_mkr}
    \min\limits_{p_k,p_r} (M_{kr})=0
\end{equation}
when $p_+=1$ and $p_-=0$, which corresponds to $p_k=p_r=1/2$. Thus, we have proven that Eq.~\ref{simplified_ineq} holds true, and subsequently, Eq.~\ref{quotient}, concluding our proof. Additionally, it is important to highlight that the inequality of Eq.~\ref{quotient} is sharp.

\section{Proofs for the illustrative examples}\label{examples}
In this section, we study the quotient $Q$ in several examples to gain an intuition about the improvement achieved with our protocol. In the previous section we saw that the minimum value of $Q_2$ was obtained for $p_k=p_r=1/2$. To analyze this case, we substitute $k=1$ and $r=2$ (and then $p_i = 0$ for $i > 3$) in Eq.~\ref{quotient} we obtain
\begin{equation}\label{sumations}
    Q_2^2=\frac{8|x_{12}|^2+\sum_{\hspace{-14pt}\substack{k, r = 1 \\ (k, r) \neq (1, 2), (2, 1)}}^{d}\hspace{-10pt}4|x_{kr}|^2}{2|x_{12}|^2}=4+\frac{\sum_{\hspace{-14pt}\substack{k, r = 1 \\ (k, r) \neq (1, 2), (2, 1)}}^{d}\hspace{-10pt}2|x_{kr}|^2}{|x_{12}|^2},
\end{equation}
where as the second term is positive, we check that the minimum value of $Q_2^2$ is actually 4 if and only if $x_{kr}=0$ when $(k,r)\neq (1,2),(2,1)$. This case corresponds to the single qubit case 
\begin{equation}\label{min_one_qubit}
    \rho=
    \begin{pmatrix}
        1/2 & 0\\
        0 & 1/2
    \end{pmatrix}
    \:
    \text{and}
    \:
    \sigma=
    \begin{pmatrix}
        1/2 & c+id\\
        c-id & 1/2
    \end{pmatrix},
\end{equation}
embedded into a larger dimension with all 0's for elements not corresponding to the ``single qubit subspace''. Notice $c$ and $d$ are real numbers chosen such that $\sigma$ is positive semi-definite, and that the diagonal elements of $\sigma$ are the same as the ones of $\rho$ to collapse the inequality from Eq.~\ref{first_inequality}. The minimum value of $Q_2=2$, was obtained for the single qubit case when considering the Frobenius norm. This result holds true for any dimension. 
\\
\\
We also conducted an analytical study using the trace norm, i.e. $Q_1$, specifically for the single qubit case. We found that the minimum quotient in the trace norm was also 2, corresponding to the same case described in Eq.~\ref{min_one_qubit}. Based on these findings and the numerical results, we intuit that the minimum lower bound of 2 applies to both the $Q_1$ and $Q_2$ and corresponds to a specific case of a single qubit.
\\
\\
To gain further insight into the quotient, we consider the maximally mixed case of $\rho=\mathds{1} / 2^q$. This corresponds to substituting $p_l=2^{-q} \: \forall l$ in Eq.~\ref{quotient}. This yields
\begin{equation}
    Q_2^2 \geq \frac{\sum_{k,r=1}^d 2^2|x_{kr}|^2}{\sum_{k,r=1}^d (2^{1-q})^2|x_{kr}|^2}= 2^{2q}.
\end{equation}
Actually, in this case, $Q_1\geq 2^q$ also holds. Thus, in the maximally mixed case, the advantage of using biomimetic copies scales with the dimension of the system $d=2^q$.
\\
\\
However, this scenario is not of real interest because if $\rho$ is maximally mixed, it commutes with $\sigma$, and the operation $e^{-i\rho t}\sigma e^{i\rho t}$ is equivalent to doing nothing. Therefore, we proceed to study a more general case
\begin{equation}
    \rho=\sum_{i=1}^d \varepsilon_i |i\rangle\langle i|,
\end{equation}
where $0\leq\varepsilon_i\leq 4/d$. This choice is made because, as noted in Ref.~\cite{Pucha_a_2016}, the eigenvalues of random density matrices distributed according to the Hilbert-Schmidt measure converge to the Marchenko-Pastur distribution, which has support between 0 and \( 4/d \). Working with the trace norm and applying the triangle inequality to the denominator and the reverse triangle inequality to the numerator
\begin{equation}
\begin{split}
    Q_1&=\frac{\norm{\;[\rho,\;\sigma]_2+2(\rho-\sigma)\;}_1}{\norm{\;[\rho,\;\sigma]_2+2\; \rho \circ\sigma-\{\rho,\sigma\}\;}_1}\\
    &\geq \frac{2\norm{\rho-\sigma}_1-\norm{[\rho,\;\sigma]_2}_1}{\norm{[\rho,\;\sigma]_2}_1+2\;\norm{\rho \circ \sigma}_1+2\;\norm{\rho\sigma}_1}.
\end{split}
\end{equation}
We can again apply the triangle inequality for the nested comutator, i.e. $\norm{[\rho,\;\sigma]_2}\leq \|\rho^2\sigma\|_1+\| \sigma\rho^2\|_1+2\;\|\rho\sigma\rho\|_1$. Using Hölder's inequality
\begin{equation}
    \norm{AB}_1\leq \norm{A}_\infty \norm{B}_1,
\end{equation}
for the different terms
\begin{itemize}
    \item $\norm{\rho\sigma}_1\leq \norm{\rho}_\infty \norm{\sigma}_1\leq 4/d$,
    \item $\| \rho^2 \sigma \|_1 \leq \|\rho^2\|_\infty \|\sigma\|_1\leq 16/d^2$,
    \item $\|\rho\sigma\rho\|_1\leq \|\rho\|_\infty \|\rho\sigma\|_1\leq 4/d \|\rho\|_\infty \|\sigma\|_1\leq 16/d^2$,
    \item $\norm{\epsilon\circ\sigma}_1=\sum_i \epsilon_i\langle i |\sigma|i\rangle\leq \sum_i \frac{4}{d}\sigma_{ii}=4/d$,
\end{itemize}
we can bound the quotient by
\begin{equation}\label{scaling_particular}
    Q_1\geq \frac{d}{8}\left(\frac{\norm{\rho-\sigma}_1-32/d^2}{1+4/d}\right).
\end{equation}
In order to further bound this expression, let us consider that $\sigma$ has rank $R<d$, and therefore we can introduce the projector $P$, such that $P\sigma P=\sigma$, $(\mathds{1}-P)\sigma(\mathds{1}-P)=0$ and $\mathrm{Rank}(P)=R$. Using this we can work on the norm of the numerator
\begin{equation}\label{term_ineq}
    \norm{\rho-\sigma}\geq \norm{P(\rho-\sigma)P}_1+\norm{(\mathds{1}-P)\rho(\mathds{1}-P)}_1.
\end{equation}
Now, we focus on the second term
\begin{equation}
    \norm{(\mathds{1}-P)\rho(\mathds{1}-P)}_1=\norm{\rho-P\rho P}_1\geq 1-\norm{P\rho P}_1
\end{equation}
where we have used the triangle inequality, along with the fact that $\|P\rho P\|_1<1$. Additionally, we have
\begin{equation}
    \norm{P\rho P}_1\leq R\;4/d.
\end{equation}
So finally, using that $\norm{P(\rho-\sigma)P}_1\geq 0$ in Eq.~\ref{term_ineq}, we get
\begin{equation}
    \norm{\rho-\sigma}_1\geq 1- R\frac{4}{d}.
\end{equation}
Therefore, based on Eq.~\ref{scaling_particular}, in this specific scenario, and when $R \ll d/4$, we observe that $Q_1$ exhibits a scaling which growths linearly with the system's dimension $d=2^q$.

\end{appendix}
\bibliography{main}

\begin{thebibliography}{55}%
\makeatletter
\providecommand \@ifxundefined [1]{%
 \@ifx{#1\undefined}
}%
\providecommand \@ifnum [1]{%
 \ifnum #1\expandafter \@firstoftwo
 \else \expandafter \@secondoftwo
 \fi
}%
\providecommand \@ifx [1]{%
 \ifx #1\expandafter \@firstoftwo
 \else \expandafter \@secondoftwo
 \fi
}%
\providecommand \natexlab [1]{#1}%
\providecommand \enquote  [1]{``#1''}%
\providecommand \bibnamefont  [1]{#1}%
\providecommand \bibfnamefont [1]{#1}%
\providecommand \citenamefont [1]{#1}%
\providecommand \href@noop [0]{\@secondoftwo}%
\providecommand \href [0]{\begingroup \@sanitize@url \@href}%
\providecommand \@href[1]{\@@startlink{#1}\@@href}%
\providecommand \@@href[1]{\endgroup#1\@@endlink}%
\providecommand \@sanitize@url [0]{\catcode `\\12\catcode `\$12\catcode
  `\&12\catcode `\#12\catcode `\^12\catcode `\_12\catcode `\%12\relax}%
\providecommand \@@startlink[1]{}%
\providecommand \@@endlink[0]{}%
\providecommand \url  [0]{\begingroup\@sanitize@url \@url }%
\providecommand \@url [1]{\endgroup\@href {#1}{\urlprefix }}%
\providecommand \urlprefix  [0]{URL }%
\providecommand \Eprint [0]{\href }%
\providecommand \doibase [0]{http://dx.doi.org/}%
\providecommand \selectlanguage [0]{\@gobble}%
\providecommand \bibinfo  [0]{\@secondoftwo}%
\providecommand \bibfield  [0]{\@secondoftwo}%
\providecommand \translation [1]{[#1]}%
\providecommand \BibitemOpen [0]{}%
\providecommand \bibitemStop [0]{}%
\providecommand \bibitemNoStop [0]{.\EOS\space}%
\providecommand \EOS [0]{\spacefactor3000\relax}%
\providecommand \BibitemShut  [1]{\csname bibitem#1\endcsname}%
\let\auto@bib@innerbib\@empty
\bibitem [{\citenamefont {Harrow}\ \emph {et~al.}(2009)\citenamefont {Harrow},
  \citenamefont {Hassidim},\ and\ \citenamefont {Lloyd}}]{HHL}%
  \BibitemOpen
  \bibfield  {author} {\bibinfo {author} {\bibfnamefont {Aram~W.}\ \bibnamefont
  {Harrow}}, \bibinfo {author} {\bibfnamefont {Avinatan}\ \bibnamefont
  {Hassidim}}, \ and\ \bibinfo {author} {\bibfnamefont {Seth}\ \bibnamefont
  {Lloyd}},\ }\bibfield  {title} {\enquote {\bibinfo {title} {Quantum algorithm
  for linear systems of equations},}\ }\href {\doibase
  10.1103/PhysRevLett.103.150502} {\bibfield  {journal} {\bibinfo  {journal}
  {Phys. Rev. Lett.}\ }\textbf {\bibinfo {volume} {103}},\ \bibinfo {pages}
  {150502} (\bibinfo {year} {2009})}\BibitemShut {NoStop}%
\bibitem [{\citenamefont {Childs}\ \emph {et~al.}(2017)\citenamefont {Childs},
  \citenamefont {Kothari},\ and\ \citenamefont {Somma}}]{HHL_Childs}%
  \BibitemOpen
  \bibfield  {author} {\bibinfo {author} {\bibfnamefont {Andrew~M.}\
  \bibnamefont {Childs}}, \bibinfo {author} {\bibfnamefont {Robin}\
  \bibnamefont {Kothari}}, \ and\ \bibinfo {author} {\bibfnamefont
  {Rolando~D.}\ \bibnamefont {Somma}},\ }\bibfield  {title} {\enquote {\bibinfo
  {title} {Quantum algorithm for systems of linear equations with exponentially
  improved dependence on precision},}\ }\href {\doibase 10.1137/16M1087072}
  {\bibfield  {journal} {\bibinfo  {journal} {SIAM Journal on Computing}\
  }\textbf {\bibinfo {volume} {46}},\ \bibinfo {pages} {1920--1950} (\bibinfo
  {year} {2017})},\ \Eprint
  {http://arxiv.org/abs/https://doi.org/10.1137/16M1087072}
  {https://doi.org/10.1137/16M1087072} \BibitemShut {NoStop}%
\bibitem [{\citenamefont {Clader}\ \emph {et~al.}(2013)\citenamefont {Clader},
  \citenamefont {Jacobs},\ and\ \citenamefont {Sprouse}}]{Preconditioned_HHL}%
  \BibitemOpen
  \bibfield  {author} {\bibinfo {author} {\bibfnamefont {B.~D.}\ \bibnamefont
  {Clader}}, \bibinfo {author} {\bibfnamefont {B.~C.}\ \bibnamefont {Jacobs}},
  \ and\ \bibinfo {author} {\bibfnamefont {C.~R.}\ \bibnamefont {Sprouse}},\
  }\bibfield  {title} {\enquote {\bibinfo {title} {Preconditioned quantum
  linear system algorithm},}\ }\href {\doibase 10.1103/PhysRevLett.110.250504}
  {\bibfield  {journal} {\bibinfo  {journal} {Phys. Rev. Lett.}\ }\textbf
  {\bibinfo {volume} {110}},\ \bibinfo {pages} {250504} (\bibinfo {year}
  {2013})}\BibitemShut {NoStop}%
\bibitem [{\citenamefont {Wiebe}\ \emph {et~al.}(2012)\citenamefont {Wiebe},
  \citenamefont {Braun},\ and\ \citenamefont {Lloyd}}]{Data_Fitting}%
  \BibitemOpen
  \bibfield  {author} {\bibinfo {author} {\bibfnamefont {Nathan}\ \bibnamefont
  {Wiebe}}, \bibinfo {author} {\bibfnamefont {Daniel}\ \bibnamefont {Braun}}, \
  and\ \bibinfo {author} {\bibfnamefont {Seth}\ \bibnamefont {Lloyd}},\
  }\bibfield  {title} {\enquote {\bibinfo {title} {Quantum algorithm for data
  fitting},}\ }\href {\doibase 10.1103/PhysRevLett.109.050505} {\bibfield
  {journal} {\bibinfo  {journal} {Phys. Rev. Lett.}\ }\textbf {\bibinfo
  {volume} {109}},\ \bibinfo {pages} {050505} (\bibinfo {year}
  {2012})}\BibitemShut {NoStop}%
\bibitem [{\citenamefont {Biamonte}\ \emph {et~al.}(2017)\citenamefont
  {Biamonte}, \citenamefont {Wittek}, \citenamefont {Pancotti}, \citenamefont
  {Rebentrost}, \citenamefont {Wiebe},\ and\ \citenamefont
  {Lloyd}}]{Biamonte_2017}%
  \BibitemOpen
  \bibfield  {author} {\bibinfo {author} {\bibfnamefont {Jacob}\ \bibnamefont
  {Biamonte}}, \bibinfo {author} {\bibfnamefont {Peter}\ \bibnamefont
  {Wittek}}, \bibinfo {author} {\bibfnamefont {Nicola}\ \bibnamefont
  {Pancotti}}, \bibinfo {author} {\bibfnamefont {Patrick}\ \bibnamefont
  {Rebentrost}}, \bibinfo {author} {\bibfnamefont {Nathan}\ \bibnamefont
  {Wiebe}}, \ and\ \bibinfo {author} {\bibfnamefont {Seth}\ \bibnamefont
  {Lloyd}},\ }\bibfield  {title} {\enquote {\bibinfo {title} {Quantum machine
  learning},}\ }\href {\doibase 10.1038/nature23474} {\bibfield  {journal}
  {\bibinfo  {journal} {Nature}\ }\textbf {\bibinfo {volume} {549}},\ \bibinfo
  {pages} {195--202} (\bibinfo {year} {2017})}\BibitemShut {NoStop}%
\bibitem [{\citenamefont {Rebentrost}\ \emph {et~al.}(2014)\citenamefont
  {Rebentrost}, \citenamefont {Mohseni},\ and\ \citenamefont
  {Lloyd}}]{Patrick_SVM_2014}%
  \BibitemOpen
  \bibfield  {author} {\bibinfo {author} {\bibfnamefont {Patrick}\ \bibnamefont
  {Rebentrost}}, \bibinfo {author} {\bibfnamefont {Masoud}\ \bibnamefont
  {Mohseni}}, \ and\ \bibinfo {author} {\bibfnamefont {Seth}\ \bibnamefont
  {Lloyd}},\ }\bibfield  {title} {\enquote {\bibinfo {title} {Quantum support
  vector machine for big data classification},}\ }\href {\doibase
  10.1103/PhysRevLett.113.130503} {\bibfield  {journal} {\bibinfo  {journal}
  {Phys. Rev. Lett.}\ }\textbf {\bibinfo {volume} {113}},\ \bibinfo {pages}
  {130503} (\bibinfo {year} {2014})}\BibitemShut {NoStop}%
\bibitem [{\citenamefont {Lloyd}\ \emph {et~al.}(2013)\citenamefont {Lloyd},
  \citenamefont {Mohseni},\ and\ \citenamefont
  {Rebentrost}}]{lloyd2013quantum}%
  \BibitemOpen
  \bibfield  {author} {\bibinfo {author} {\bibfnamefont {Seth}\ \bibnamefont
  {Lloyd}}, \bibinfo {author} {\bibfnamefont {Masoud}\ \bibnamefont {Mohseni}},
  \ and\ \bibinfo {author} {\bibfnamefont {Patrick}\ \bibnamefont
  {Rebentrost}},\ }\href@noop {} {\enquote {\bibinfo {title} {Quantum
  algorithms for supervised and unsupervised machine learning},}\ } (\bibinfo
  {year} {2013}),\ \Eprint {http://arxiv.org/abs/1307.0411} {arXiv:1307.0411
  [quant-ph]} \BibitemShut {NoStop}%
\bibitem [{\citenamefont {Huang}\ \emph {et~al.}(2021)\citenamefont {Huang},
  \citenamefont {Broughton}, \citenamefont {Mohseni}, \citenamefont {Babbush},
  \citenamefont {Boixo}, \citenamefont {Neven},\ and\ \citenamefont
  {McClean}}]{Robert_power_data}%
  \BibitemOpen
  \bibfield  {author} {\bibinfo {author} {\bibfnamefont {Hsin-Yuan}\
  \bibnamefont {Huang}}, \bibinfo {author} {\bibfnamefont {Michael}\
  \bibnamefont {Broughton}}, \bibinfo {author} {\bibfnamefont {Masoud}\
  \bibnamefont {Mohseni}}, \bibinfo {author} {\bibfnamefont {Ryan}\
  \bibnamefont {Babbush}}, \bibinfo {author} {\bibfnamefont {Sergio}\
  \bibnamefont {Boixo}}, \bibinfo {author} {\bibfnamefont {Hartmut}\
  \bibnamefont {Neven}}, \ and\ \bibinfo {author} {\bibfnamefont {Jarrod~R.}\
  \bibnamefont {McClean}},\ }\bibfield  {title} {\enquote {\bibinfo {title}
  {Power of data in quantum machine learning},}\ }\href {\doibase
  10.1038/s41467-021-22539-9} {\bibfield  {journal} {\bibinfo  {journal}
  {Nature Communications}\ }\textbf {\bibinfo {volume} {12}},\ \bibinfo {pages}
  {2631} (\bibinfo {year} {2021})}\BibitemShut {NoStop}%
\bibitem [{\citenamefont {Schuld}\ and\ \citenamefont
  {Killoran}(2019)}]{Schuld_2019}%
  \BibitemOpen
  \bibfield  {author} {\bibinfo {author} {\bibfnamefont {Maria}\ \bibnamefont
  {Schuld}}\ and\ \bibinfo {author} {\bibfnamefont {Nathan}\ \bibnamefont
  {Killoran}},\ }\bibfield  {title} {\enquote {\bibinfo {title} {Quantum
  machine learning in feature hilbert spaces},}\ }\href {\doibase
  10.1103/physrevlett.122.040504} {\bibfield  {journal} {\bibinfo  {journal}
  {Physical Review Letters}\ }\textbf {\bibinfo {volume} {122}} (\bibinfo
  {year} {2019}),\ 10.1103/physrevlett.122.040504}\BibitemShut {NoStop}%
\bibitem [{\citenamefont {Lloyd}\ \emph {et~al.}(2014)\citenamefont {Lloyd},
  \citenamefont {Mohseni},\ and\ \citenamefont {Rebentrost}}]{Patrick_PCA}%
  \BibitemOpen
  \bibfield  {author} {\bibinfo {author} {\bibfnamefont {Seth}\ \bibnamefont
  {Lloyd}}, \bibinfo {author} {\bibfnamefont {Masoud}\ \bibnamefont {Mohseni}},
  \ and\ \bibinfo {author} {\bibfnamefont {Patrick}\ \bibnamefont
  {Rebentrost}},\ }\bibfield  {title} {\enquote {\bibinfo {title} {Quantum
  principal component analysis},}\ }\href {\doibase 10.1038/nphys3029}
  {\bibfield  {journal} {\bibinfo  {journal} {Nature Physics}\ }\textbf
  {\bibinfo {volume} {10}},\ \bibinfo {pages} {631--633} (\bibinfo {year}
  {2014})}\BibitemShut {NoStop}%
\bibitem [{\citenamefont {Schuld}\ \emph {et~al.}(2021)\citenamefont {Schuld},
  \citenamefont {Sweke},\ and\ \citenamefont {Meyer}}]{Encoding_variational}%
  \BibitemOpen
  \bibfield  {author} {\bibinfo {author} {\bibfnamefont {Maria}\ \bibnamefont
  {Schuld}}, \bibinfo {author} {\bibfnamefont {Ryan}\ \bibnamefont {Sweke}}, \
  and\ \bibinfo {author} {\bibfnamefont {Johannes~Jakob}\ \bibnamefont
  {Meyer}},\ }\bibfield  {title} {\enquote {\bibinfo {title} {Effect of data
  encoding on the expressive power of variational quantum-machine-learning
  models},}\ }\href {\doibase 10.1103/PhysRevA.103.032430} {\bibfield
  {journal} {\bibinfo  {journal} {Phys. Rev. A}\ }\textbf {\bibinfo {volume}
  {103}},\ \bibinfo {pages} {032430} (\bibinfo {year} {2021})}\BibitemShut
  {NoStop}%
\bibitem [{\citenamefont {Lloyd}\ \emph {et~al.}(2020)\citenamefont {Lloyd},
  \citenamefont {Schuld}, \citenamefont {Ijaz}, \citenamefont {Izaac},\ and\
  \citenamefont {Killoran}}]{lloyd2020quantum}%
  \BibitemOpen
  \bibfield  {author} {\bibinfo {author} {\bibfnamefont {Seth}\ \bibnamefont
  {Lloyd}}, \bibinfo {author} {\bibfnamefont {Maria}\ \bibnamefont {Schuld}},
  \bibinfo {author} {\bibfnamefont {Aroosa}\ \bibnamefont {Ijaz}}, \bibinfo
  {author} {\bibfnamefont {Josh}\ \bibnamefont {Izaac}}, \ and\ \bibinfo
  {author} {\bibfnamefont {Nathan}\ \bibnamefont {Killoran}},\ }\href@noop {}
  {\enquote {\bibinfo {title} {Quantum embeddings for machine learning},}\ }
  (\bibinfo {year} {2020}),\ \Eprint {http://arxiv.org/abs/2001.03622}
  {arXiv:2001.03622 [quant-ph]} \BibitemShut {NoStop}%
\bibitem [{\citenamefont {Schuld}\ and\ \citenamefont
  {Petruccione}(2021)}]{Schuld2021}%
  \BibitemOpen
  \bibfield  {author} {\bibinfo {author} {\bibfnamefont {Maria}\ \bibnamefont
  {Schuld}}\ and\ \bibinfo {author} {\bibfnamefont {Francesco}\ \bibnamefont
  {Petruccione}},\ }\enquote {\bibinfo {title} {Quantum models as kernel
  methods},}\ in\ \href {\doibase 10.1007/978-3-030-83098-4_6} {\emph {\bibinfo
  {booktitle} {Machine Learning with Quantum Computers}}}\ (\bibinfo
  {publisher} {Springer International Publishing},\ \bibinfo {address} {Cham},\
  \bibinfo {year} {2021})\ pp.\ \bibinfo {pages} {217--245}\BibitemShut
  {NoStop}%
\bibitem [{\citenamefont {Rebentrost}\ \emph {et~al.}(2018)\citenamefont
  {Rebentrost}, \citenamefont {Gupt},\ and\ \citenamefont
  {Bromley}}]{Patrick_Options}%
  \BibitemOpen
  \bibfield  {author} {\bibinfo {author} {\bibfnamefont {Patrick}\ \bibnamefont
  {Rebentrost}}, \bibinfo {author} {\bibfnamefont {Brajesh}\ \bibnamefont
  {Gupt}}, \ and\ \bibinfo {author} {\bibfnamefont {Thomas~R.}\ \bibnamefont
  {Bromley}},\ }\bibfield  {title} {\enquote {\bibinfo {title} {{Quantum
  computational finance: Monte Carlo pricing of financial derivatives}},}\
  }\href {\doibase 10.1103/PhysRevA.98.022321} {\bibfield  {journal} {\bibinfo
  {journal} {Phys. Rev. A}\ }\textbf {\bibinfo {volume} {98}},\ \bibinfo
  {pages} {022321} (\bibinfo {year} {2018})}\BibitemShut {NoStop}%
\bibitem [{\citenamefont {Marin-Sanchez}\ \emph {et~al.}(2021)\citenamefont
  {Marin-Sanchez}, \citenamefont {Gonzalez-Conde},\ and\ \citenamefont
  {Sanz}}]{marinsanchez2021quantum}%
  \BibitemOpen
  \bibfield  {author} {\bibinfo {author} {\bibfnamefont {Gabriel}\ \bibnamefont
  {Marin-Sanchez}}, \bibinfo {author} {\bibfnamefont {Javier}\ \bibnamefont
  {Gonzalez-Conde}}, \ and\ \bibinfo {author} {\bibfnamefont {Mikel}\
  \bibnamefont {Sanz}},\ }\href@noop {} {\enquote {\bibinfo {title} {Quantum
  algorithms for approximate function loading},}\ } (\bibinfo {year} {2021}),\
  \Eprint {http://arxiv.org/abs/2111.07933} {arXiv:2111.07933 [quant-ph]}
  \BibitemShut {NoStop}%
\bibitem [{\citenamefont {Gonzalez-Conde}\ \emph {et~al.}(2024)\citenamefont
  {Gonzalez-Conde}, \citenamefont {Watts}, \citenamefont {Rodriguez-Grasa},\
  and\ \citenamefont {Sanz}}]{gonzalezconde2023efficient}%
  \BibitemOpen
  \bibfield  {author} {\bibinfo {author} {\bibfnamefont {Javier}\ \bibnamefont
  {Gonzalez-Conde}}, \bibinfo {author} {\bibfnamefont {Thomas~W.}\ \bibnamefont
  {Watts}}, \bibinfo {author} {\bibfnamefont {Pablo}\ \bibnamefont
  {Rodriguez-Grasa}}, \ and\ \bibinfo {author} {\bibfnamefont {Mikel}\
  \bibnamefont {Sanz}},\ }\bibfield  {title} {\enquote {\bibinfo {title}
  {Efficient quantum amplitude encoding of polynomial functions},}\ }\href
  {\doibase 10.22331/q-2024-03-21-1297} {\bibfield  {journal} {\bibinfo
  {journal} {Quantum}\ }\textbf {\bibinfo {volume} {8}},\ \bibinfo {pages}
  {1297} (\bibinfo {year} {2024})}\BibitemShut {NoStop}%
\bibitem [{\citenamefont {Gonzalez-Conde}\ \emph {et~al.}(2022)\citenamefont
  {Gonzalez-Conde}, \citenamefont {Ángel Rodríguez-Rozas}, \citenamefont
  {Solano},\ and\ \citenamefont {Sanz}}]{gonzalezconde2022simulating}%
  \BibitemOpen
  \bibfield  {author} {\bibinfo {author} {\bibfnamefont {Javier}\ \bibnamefont
  {Gonzalez-Conde}}, \bibinfo {author} {\bibnamefont {Ángel
  Rodríguez-Rozas}}, \bibinfo {author} {\bibfnamefont {Enrique}\ \bibnamefont
  {Solano}}, \ and\ \bibinfo {author} {\bibfnamefont {Mikel}\ \bibnamefont
  {Sanz}},\ }\href@noop {} {\enquote {\bibinfo {title} {Simulating option price
  dynamics with exponential quantum speedup},}\ } (\bibinfo {year} {2022}),\
  \Eprint {http://arxiv.org/abs/2101.04023} {arXiv:2101.04023 [quant-ph]}
  \BibitemShut {NoStop}%
\bibitem [{\citenamefont {Martin}\ \emph {et~al.}(2021)\citenamefont {Martin},
  \citenamefont {Candelas}, \citenamefont {Rodr{\'{\i}}guez-Rozas},
  \citenamefont {Mart{\'{\i}}n-Guerrero}, \citenamefont {Chen}, \citenamefont
  {Lamata}, \citenamefont {Or{\'{u}}s}, \citenamefont {Solano},\ and\
  \citenamefont {Sanz}}]{Martin_2021}%
  \BibitemOpen
  \bibfield  {author} {\bibinfo {author} {\bibfnamefont {Ana}\ \bibnamefont
  {Martin}}, \bibinfo {author} {\bibfnamefont {Bruno}\ \bibnamefont
  {Candelas}}, \bibinfo {author} {\bibfnamefont {{\'{A} }ngel}\ \bibnamefont
  {Rodr{\'{\i}}guez-Rozas}}, \bibinfo {author} {\bibfnamefont {Jos{\'{e}}~D.}\
  \bibnamefont {Mart{\'{\i}}n-Guerrero}}, \bibinfo {author} {\bibfnamefont
  {Xi}~\bibnamefont {Chen}}, \bibinfo {author} {\bibfnamefont {Lucas}\
  \bibnamefont {Lamata}}, \bibinfo {author} {\bibfnamefont {Rom{\'{a}}n}\
  \bibnamefont {Or{\'{u}}s}}, \bibinfo {author} {\bibfnamefont {Enrique}\
  \bibnamefont {Solano}}, \ and\ \bibinfo {author} {\bibfnamefont {Mikel}\
  \bibnamefont {Sanz}},\ }\bibfield  {title} {\enquote {\bibinfo {title}
  {Toward pricing financial derivatives with an {IBM} quantum computer},}\
  }\href {\doibase 10.1103/physrevresearch.3.013167} {\bibfield  {journal}
  {\bibinfo  {journal} {Physical Review Research}\ }\textbf {\bibinfo {volume}
  {3}} (\bibinfo {year} {2021}),\ 10.1103/physrevresearch.3.013167}\BibitemShut
  {NoStop}%
\bibitem [{\citenamefont {Martin}\ \emph {et~al.}(2023)\citenamefont {Martin},
  \citenamefont {Ibarrondo},\ and\ \citenamefont
  {Sanz}}]{martin2022digitalanalog}%
  \BibitemOpen
  \bibfield  {author} {\bibinfo {author} {\bibfnamefont {Ana}\ \bibnamefont
  {Martin}}, \bibinfo {author} {\bibfnamefont {Ruben}\ \bibnamefont
  {Ibarrondo}}, \ and\ \bibinfo {author} {\bibfnamefont {Mikel}\ \bibnamefont
  {Sanz}},\ }\bibfield  {title} {\enquote {\bibinfo {title} {Digital-analog
  co-design of the harrow-hassidim-lloyd algorithm},}\ }\href {\doibase
  10.1103/physrevapplied.19.064056} {\bibfield  {journal} {\bibinfo  {journal}
  {Physical Review Applied}\ }\textbf {\bibinfo {volume} {19}} (\bibinfo {year}
  {2023}),\ 10.1103/physrevapplied.19.064056}\BibitemShut {NoStop}%
\bibitem [{\citenamefont {Kimmel}\ \emph {et~al.}(2017)\citenamefont {Kimmel},
  \citenamefont {Lin}, \citenamefont {Low}, \citenamefont {Ozols},\ and\
  \citenamefont {Yoder}}]{Kimmel_2017}%
  \BibitemOpen
  \bibfield  {author} {\bibinfo {author} {\bibfnamefont {Shelby}\ \bibnamefont
  {Kimmel}}, \bibinfo {author} {\bibfnamefont {Cedric Yen-Yu}\ \bibnamefont
  {Lin}}, \bibinfo {author} {\bibfnamefont {Guang~Hao}\ \bibnamefont {Low}},
  \bibinfo {author} {\bibfnamefont {Maris}\ \bibnamefont {Ozols}}, \ and\
  \bibinfo {author} {\bibfnamefont {Theodore~J.}\ \bibnamefont {Yoder}},\
  }\bibfield  {title} {\enquote {\bibinfo {title} {Hamiltonian simulation with
  optimal sample complexity},}\ }\href {\doibase 10.1038/s41534-017-0013-7}
  {\bibfield  {journal} {\bibinfo  {journal} {npj Quantum Information}\
  }\textbf {\bibinfo {volume} {3}} (\bibinfo {year} {2017}),\
  10.1038/s41534-017-0013-7}\BibitemShut {NoStop}%
\bibitem [{\citenamefont {Trotter}(1959)}]{Trotter_1959}%
  \BibitemOpen
  \bibfield  {author} {\bibinfo {author} {\bibfnamefont {Hale~F.}\ \bibnamefont
  {Trotter}},\ }\bibfield  {title} {\enquote {\bibinfo {title} {On the product
  of semi-groups of operators},}\ \ }(\bibinfo {year} {1959})\BibitemShut
  {NoStop}%
\bibitem [{\citenamefont {Suzuki}(1976)}]{Suzuki_1976}%
  \BibitemOpen
  \bibfield  {author} {\bibinfo {author} {\bibfnamefont {M.}~\bibnamefont
  {Suzuki}},\ }\bibfield  {title} {\enquote {\bibinfo {title} {{Generalized
  Trotter's Formula and Systematic Approximants of Exponential Operators and
  Inner Derivations with Applications to Many Body Problems}},}\ }\href
  {\doibase 10.1007/BF01609348} {\bibfield  {journal} {\bibinfo  {journal}
  {Commun. Math. Phys.}\ }\textbf {\bibinfo {volume} {51}},\ \bibinfo {pages}
  {183--190} (\bibinfo {year} {1976})}\BibitemShut {NoStop}%
\bibitem [{\citenamefont {Barends}\ \emph {et~al.}(2016)\citenamefont
  {Barends}, \citenamefont {Shabani}, \citenamefont {Lamata}, \citenamefont
  {Kelly}, \citenamefont {Mezzacapo}, \citenamefont {Heras}, \citenamefont
  {Babbush}, \citenamefont {Fowler}, \citenamefont {Campbell}, \citenamefont
  {Chen}, \citenamefont {Chen}, \citenamefont {Chiaro}, \citenamefont
  {Dunsworth}, \citenamefont {Jeffrey}, \citenamefont {Lucero}, \citenamefont
  {Megrant}, \citenamefont {Mutus}, \citenamefont {Neeley}, \citenamefont
  {Neill}, \citenamefont {O'Malley}, \citenamefont {Quintana}, \citenamefont
  {Solano}, \citenamefont {White}, \citenamefont {Wenner}, \citenamefont
  {Vainsencher}, \citenamefont {Sank}, \citenamefont {Roushan}, \citenamefont
  {Neven},\ and\ \citenamefont {Martinis}}]{Barends_2016}%
  \BibitemOpen
  \bibfield  {author} {\bibinfo {author} {\bibfnamefont {Rami}\ \bibnamefont
  {Barends}}, \bibinfo {author} {\bibfnamefont {Alireza}\ \bibnamefont
  {Shabani}}, \bibinfo {author} {\bibfnamefont {Lucas}\ \bibnamefont {Lamata}},
  \bibinfo {author} {\bibfnamefont {Julian}\ \bibnamefont {Kelly}}, \bibinfo
  {author} {\bibfnamefont {Antonio}\ \bibnamefont {Mezzacapo}}, \bibinfo
  {author} {\bibfnamefont {Urtzi~Las}\ \bibnamefont {Heras}}, \bibinfo {author}
  {\bibfnamefont {Ryan}\ \bibnamefont {Babbush}}, \bibinfo {author}
  {\bibfnamefont {Austin}\ \bibnamefont {Fowler}}, \bibinfo {author}
  {\bibfnamefont {Brooks}\ \bibnamefont {Campbell}}, \bibinfo {author}
  {\bibfnamefont {Yu}~\bibnamefont {Chen}}, \bibinfo {author} {\bibfnamefont
  {Zijun}\ \bibnamefont {Chen}}, \bibinfo {author} {\bibfnamefont {Ben}\
  \bibnamefont {Chiaro}}, \bibinfo {author} {\bibfnamefont {Andrew}\
  \bibnamefont {Dunsworth}}, \bibinfo {author} {\bibfnamefont {Evan}\
  \bibnamefont {Jeffrey}}, \bibinfo {author} {\bibfnamefont {Erik}\
  \bibnamefont {Lucero}}, \bibinfo {author} {\bibfnamefont {Anthony}\
  \bibnamefont {Megrant}}, \bibinfo {author} {\bibfnamefont {Josh}\
  \bibnamefont {Mutus}}, \bibinfo {author} {\bibfnamefont {Matthew}\
  \bibnamefont {Neeley}}, \bibinfo {author} {\bibfnamefont {Charles}\
  \bibnamefont {Neill}}, \bibinfo {author} {\bibfnamefont {Peter}\ \bibnamefont
  {O'Malley}}, \bibinfo {author} {\bibfnamefont {Chris}\ \bibnamefont
  {Quintana}}, \bibinfo {author} {\bibfnamefont {Enrique}\ \bibnamefont
  {Solano}}, \bibinfo {author} {\bibfnamefont {Ted}\ \bibnamefont {White}},
  \bibinfo {author} {\bibfnamefont {Jim}\ \bibnamefont {Wenner}}, \bibinfo
  {author} {\bibfnamefont {Amit}\ \bibnamefont {Vainsencher}}, \bibinfo
  {author} {\bibfnamefont {Daniel}\ \bibnamefont {Sank}}, \bibinfo {author}
  {\bibfnamefont {Pedram}\ \bibnamefont {Roushan}}, \bibinfo {author}
  {\bibfnamefont {Hartmut}\ \bibnamefont {Neven}}, \ and\ \bibinfo {author}
  {\bibfnamefont {John}\ \bibnamefont {Martinis}},\ }\bibfield  {title}
  {\enquote {\bibinfo {title} {Digitized adiabatic quantum computing with a
  superconducting circuit},}\ }\href
  {http://www.nature.com/nature/journal/v534/n7606/full/nature17658.html}
  {\bibfield  {journal} {\bibinfo  {journal} {Nature}\ }\textbf {\bibinfo
  {volume} {534}},\ \bibinfo {pages} {222--226} (\bibinfo {year}
  {2016})}\BibitemShut {NoStop}%
\bibitem [{\citenamefont {Layden}(2022)}]{Layden2022}%
  \BibitemOpen
  \bibfield  {author} {\bibinfo {author} {\bibfnamefont {David}\ \bibnamefont
  {Layden}},\ }\bibfield  {title} {\enquote {\bibinfo {title} {First-order
  trotter error from a second-order perspective},}\ }\href {\doibase
  10.1103/PhysRevLett.128.210501} {\bibfield  {journal} {\bibinfo  {journal}
  {Physical Review Letters}\ }\textbf {\bibinfo {volume} {128}} (\bibinfo
  {year} {2022}),\ 10.1103/PhysRevLett.128.210501}\BibitemShut {NoStop}%
\bibitem [{\citenamefont {Childs}\ \emph {et~al.}(2021)\citenamefont {Childs},
  \citenamefont {Su}, \citenamefont {Tran}, \citenamefont {Wiebe},\ and\
  \citenamefont {Zhu}}]{Childs_2021}%
  \BibitemOpen
  \bibfield  {author} {\bibinfo {author} {\bibfnamefont {Andrew~M.}\
  \bibnamefont {Childs}}, \bibinfo {author} {\bibfnamefont {Yuan}\ \bibnamefont
  {Su}}, \bibinfo {author} {\bibfnamefont {Minh~C.}\ \bibnamefont {Tran}},
  \bibinfo {author} {\bibfnamefont {Nathan}\ \bibnamefont {Wiebe}}, \ and\
  \bibinfo {author} {\bibfnamefont {Shuchen}\ \bibnamefont {Zhu}},\ }\bibfield
  {title} {\enquote {\bibinfo {title} {Theory of trotter error with commutator
  scaling},}\ }\href {\doibase 10.1103/PhysRevX.11.011020} {\bibfield
  {journal} {\bibinfo  {journal} {Phys. Rev. X}\ }\textbf {\bibinfo {volume}
  {11}},\ \bibinfo {pages} {011020} (\bibinfo {year} {2021})}\BibitemShut
  {NoStop}%
\bibitem [{\citenamefont {Campbell}(2019)}]{Campbell_2019}%
  \BibitemOpen
  \bibfield  {author} {\bibinfo {author} {\bibfnamefont {Earl}\ \bibnamefont
  {Campbell}},\ }\bibfield  {title} {\enquote {\bibinfo {title} {Random
  compiler for fast hamiltonian simulation},}\ }\href {\doibase
  10.1103/physrevlett.123.070503} {\bibfield  {journal} {\bibinfo  {journal}
  {Physical Review Letters}\ }\textbf {\bibinfo {volume} {123}} (\bibinfo
  {year} {2019}),\ 10.1103/physrevlett.123.070503}\BibitemShut {NoStop}%
\bibitem [{\citenamefont {Low}\ and\ \citenamefont {Chuang}(2019)}]{Low_2019}%
  \BibitemOpen
  \bibfield  {author} {\bibinfo {author} {\bibfnamefont {Guang~Hao}\
  \bibnamefont {Low}}\ and\ \bibinfo {author} {\bibfnamefont {Isaac~L.}\
  \bibnamefont {Chuang}},\ }\bibfield  {title} {\enquote {\bibinfo {title}
  {Hamiltonian simulation by qubitization},}\ }\href {\doibase
  10.22331/q-2019-07-12-163} {\bibfield  {journal} {\bibinfo  {journal}
  {Quantum}\ }\textbf {\bibinfo {volume} {3}},\ \bibinfo {pages} {163}
  (\bibinfo {year} {2019})}\BibitemShut {NoStop}%
\bibitem [{\citenamefont {Berry}\ \emph
  {et~al.}(2015{\natexlab{a}})\citenamefont {Berry}, \citenamefont {Childs},\
  and\ \citenamefont {Kothari}}]{Berry_2015}%
  \BibitemOpen
  \bibfield  {author} {\bibinfo {author} {\bibfnamefont {Dominic~W.}\
  \bibnamefont {Berry}}, \bibinfo {author} {\bibfnamefont {Andrew~M.}\
  \bibnamefont {Childs}}, \ and\ \bibinfo {author} {\bibfnamefont {Robin}\
  \bibnamefont {Kothari}},\ }\bibfield  {title} {\enquote {\bibinfo {title}
  {Hamiltonian simulation with nearly optimal dependence on all parameters},}\
  }in\ \href {\doibase 10.1109/focs.2015.54} {\emph {\bibinfo {booktitle} {2015
  {IEEE} 56th Annual Symposium on Foundations of Computer Science}}}\ (\bibinfo
   {publisher} {{IEEE}},\ \bibinfo {year} {2015})\BibitemShut {NoStop}%
\bibitem [{\citenamefont {Childs}\ and\ \citenamefont
  {Wiebe}(2012)}]{Childs2012}%
  \BibitemOpen
  \bibfield  {author} {\bibinfo {author} {\bibfnamefont {Andrew~M.}\
  \bibnamefont {Childs}}\ and\ \bibinfo {author} {\bibfnamefont {Nathan}\
  \bibnamefont {Wiebe}},\ }\bibfield  {title} {\enquote {\bibinfo {title}
  {Hamiltonian simulation using linear combinations of unitary operations},}\
  }\href@noop {} {\bibfield  {journal} {\bibinfo  {journal} {Quantum Info.
  Comput.}\ }\textbf {\bibinfo {volume} {12}},\ \bibinfo {pages} {901–924}
  (\bibinfo {year} {2012})}\BibitemShut {NoStop}%
\bibitem [{\citenamefont {Berry}\ \emph {et~al.}(2014)\citenamefont {Berry},
  \citenamefont {Childs}, \citenamefont {Cleve}, \citenamefont {Kothari},\ and\
  \citenamefont {Somma}}]{Berry_2014}%
  \BibitemOpen
  \bibfield  {author} {\bibinfo {author} {\bibfnamefont {Dominic~W.}\
  \bibnamefont {Berry}}, \bibinfo {author} {\bibfnamefont {Andrew~M.}\
  \bibnamefont {Childs}}, \bibinfo {author} {\bibfnamefont {Richard}\
  \bibnamefont {Cleve}}, \bibinfo {author} {\bibfnamefont {Robin}\ \bibnamefont
  {Kothari}}, \ and\ \bibinfo {author} {\bibfnamefont {Rolando~D.}\
  \bibnamefont {Somma}},\ }\bibfield  {title} {\enquote {\bibinfo {title}
  {Exponential improvement in precision for simulating sparse hamiltonians},}\
  }in\ \href {\doibase 10.1145/2591796.2591854} {\emph {\bibinfo {booktitle}
  {Proceedings of the forty-sixth annual {ACM} symposium on Theory of
  computing}}}\ (\bibinfo  {publisher} {{ACM}},\ \bibinfo {year}
  {2014})\BibitemShut {NoStop}%
\bibitem [{\citenamefont {Berry}\ \emph
  {et~al.}(2015{\natexlab{b}})\citenamefont {Berry}, \citenamefont {Childs},
  \citenamefont {Cleve}, \citenamefont {Kothari},\ and\ \citenamefont
  {Somma}}]{Berry_2015_2}%
  \BibitemOpen
  \bibfield  {author} {\bibinfo {author} {\bibfnamefont {Dominic~W.}\
  \bibnamefont {Berry}}, \bibinfo {author} {\bibfnamefont {Andrew~M.}\
  \bibnamefont {Childs}}, \bibinfo {author} {\bibfnamefont {Richard}\
  \bibnamefont {Cleve}}, \bibinfo {author} {\bibfnamefont {Robin}\ \bibnamefont
  {Kothari}}, \ and\ \bibinfo {author} {\bibfnamefont {Rolando~D.}\
  \bibnamefont {Somma}},\ }\bibfield  {title} {\enquote {\bibinfo {title}
  {Simulating hamiltonian dynamics with a truncated taylor series},}\ }\href
  {\doibase 10.1103/PhysRevLett.114.090502} {\bibfield  {journal} {\bibinfo
  {journal} {Phys. Rev. Lett.}\ }\textbf {\bibinfo {volume} {114}},\ \bibinfo
  {pages} {090502} (\bibinfo {year} {2015}{\natexlab{b}})}\BibitemShut
  {NoStop}%
\bibitem [{\citenamefont {K\"okc\"u}\ \emph {et~al.}(2022)\citenamefont
  {K\"okc\"u}, \citenamefont {Steckmann}, \citenamefont {Wang}, \citenamefont
  {Freericks}, \citenamefont {Dumitrescu},\ and\ \citenamefont
  {Kemper}}]{Efekan_2022}%
  \BibitemOpen
  \bibfield  {author} {\bibinfo {author} {\bibfnamefont {Efekan}\ \bibnamefont
  {K\"okc\"u}}, \bibinfo {author} {\bibfnamefont {Thomas}\ \bibnamefont
  {Steckmann}}, \bibinfo {author} {\bibfnamefont {Yan}\ \bibnamefont {Wang}},
  \bibinfo {author} {\bibfnamefont {J.~K.}\ \bibnamefont {Freericks}}, \bibinfo
  {author} {\bibfnamefont {Eugene~F.}\ \bibnamefont {Dumitrescu}}, \ and\
  \bibinfo {author} {\bibfnamefont {Alexander~F.}\ \bibnamefont {Kemper}},\
  }\bibfield  {title} {\enquote {\bibinfo {title} {Fixed depth hamiltonian
  simulation via cartan decomposition},}\ }\href {\doibase
  10.1103/PhysRevLett.129.070501} {\bibfield  {journal} {\bibinfo  {journal}
  {Phys. Rev. Lett.}\ }\textbf {\bibinfo {volume} {129}},\ \bibinfo {pages}
  {070501} (\bibinfo {year} {2022})}\BibitemShut {NoStop}%
\bibitem [{\citenamefont {Low}\ and\ \citenamefont {Chuang}(2017)}]{Low_2017}%
  \BibitemOpen
  \bibfield  {author} {\bibinfo {author} {\bibfnamefont {Guang~Hao}\
  \bibnamefont {Low}}\ and\ \bibinfo {author} {\bibfnamefont {Isaac~L.}\
  \bibnamefont {Chuang}},\ }\bibfield  {title} {\enquote {\bibinfo {title}
  {Optimal hamiltonian simulation by quantum signal processing},}\ }\href
  {\doibase 10.1103/PhysRevLett.118.010501} {\bibfield  {journal} {\bibinfo
  {journal} {Phys. Rev. Lett.}\ }\textbf {\bibinfo {volume} {118}},\ \bibinfo
  {pages} {010501} (\bibinfo {year} {2017})}\BibitemShut {NoStop}%
\bibitem [{\citenamefont {C{\^{\i}}rstoiu}\ \emph {et~al.}(2020)\citenamefont
  {C{\^{\i}}rstoiu}, \citenamefont {Holmes}, \citenamefont {Iosue},
  \citenamefont {Cincio}, \citenamefont {Coles},\ and\ \citenamefont
  {Sornborger}}]{C_rstoiu_2020}%
  \BibitemOpen
  \bibfield  {author} {\bibinfo {author} {\bibfnamefont {Cristina}\
  \bibnamefont {C{\^{\i}}rstoiu}}, \bibinfo {author} {\bibfnamefont {Zoë}\
  \bibnamefont {Holmes}}, \bibinfo {author} {\bibfnamefont {Joseph}\
  \bibnamefont {Iosue}}, \bibinfo {author} {\bibfnamefont {Lukasz}\
  \bibnamefont {Cincio}}, \bibinfo {author} {\bibfnamefont {Patrick~J.}\
  \bibnamefont {Coles}}, \ and\ \bibinfo {author} {\bibfnamefont {Andrew}\
  \bibnamefont {Sornborger}},\ }\bibfield  {title} {\enquote {\bibinfo {title}
  {Variational fast forwarding for quantum simulation beyond the coherence
  time},}\ }\href {\doibase 10.1038/s41534-020-00302-0} {\bibfield  {journal}
  {\bibinfo  {journal} {npj Quantum Information}\ }\textbf {\bibinfo {volume}
  {6}} (\bibinfo {year} {2020}),\ 10.1038/s41534-020-00302-0}\BibitemShut
  {NoStop}%
\bibitem [{\citenamefont {Dong}\ \emph {et~al.}(2022)\citenamefont {Dong},
  \citenamefont {Whaley},\ and\ \citenamefont {Lin}}]{Dong_2021}%
  \BibitemOpen
  \bibfield  {author} {\bibinfo {author} {\bibfnamefont {Yulong}\ \bibnamefont
  {Dong}}, \bibinfo {author} {\bibfnamefont {K.~Birgitta}\ \bibnamefont
  {Whaley}}, \ and\ \bibinfo {author} {\bibfnamefont {Lin}\ \bibnamefont
  {Lin}},\ }\bibfield  {title} {\enquote {\bibinfo {title} {A quantum
  hamiltonian simulation benchmark},}\ }\href {\doibase
  10.1038/s41534-022-00636-x} {\bibfield  {journal} {\bibinfo  {journal} {npj
  Quantum Information}\ }\textbf {\bibinfo {volume} {8}} (\bibinfo {year}
  {2022}),\ 10.1038/s41534-022-00636-x}\BibitemShut {NoStop}%
\bibitem [{\citenamefont {Gilyén}\ and\ \citenamefont
  {Poremba}(2022)}]{gilyén2022improved}%
  \BibitemOpen
  \bibfield  {author} {\bibinfo {author} {\bibfnamefont {András}\ \bibnamefont
  {Gilyén}}\ and\ \bibinfo {author} {\bibfnamefont {Alexander}\ \bibnamefont
  {Poremba}},\ }\href@noop {} {\enquote {\bibinfo {title} {Improved quantum
  algorithms for fidelity estimation},}\ } (\bibinfo {year} {2022}),\ \Eprint
  {http://arxiv.org/abs/2203.15993} {arXiv:2203.15993 [quant-ph]} \BibitemShut
  {NoStop}%
\bibitem [{\citenamefont {Marvian}\ and\ \citenamefont
  {Lloyd}(2016)}]{marvian2016universal}%
  \BibitemOpen
  \bibfield  {author} {\bibinfo {author} {\bibfnamefont {Iman}\ \bibnamefont
  {Marvian}}\ and\ \bibinfo {author} {\bibfnamefont {Seth}\ \bibnamefont
  {Lloyd}},\ }\href@noop {} {\enquote {\bibinfo {title} {Universal quantum
  emulator},}\ } (\bibinfo {year} {2016}),\ \Eprint
  {http://arxiv.org/abs/1606.02734} {arXiv:1606.02734 [quant-ph]} \BibitemShut
  {NoStop}%
\bibitem [{\citenamefont {Huggins}\ and\ \citenamefont
  {McClean}(2023)}]{huggins2023accelerating}%
  \BibitemOpen
  \bibfield  {author} {\bibinfo {author} {\bibfnamefont {William~J.}\
  \bibnamefont {Huggins}}\ and\ \bibinfo {author} {\bibfnamefont {Jarrod~R.}\
  \bibnamefont {McClean}},\ }\href@noop {} {\enquote {\bibinfo {title}
  {Accelerating quantum algorithms with precomputation},}\ } (\bibinfo {year}
  {2023}),\ \Eprint {http://arxiv.org/abs/2305.09638} {arXiv:2305.09638
  [quant-ph]} \BibitemShut {NoStop}%
\bibitem [{\citenamefont {Berry}\ \emph {et~al.}(2006)\citenamefont {Berry},
  \citenamefont {Ahokas}, \citenamefont {Cleve},\ and\ \citenamefont
  {Sanders}}]{Berry_2006}%
  \BibitemOpen
  \bibfield  {author} {\bibinfo {author} {\bibfnamefont {Dominic~W.}\
  \bibnamefont {Berry}}, \bibinfo {author} {\bibfnamefont {Graeme}\
  \bibnamefont {Ahokas}}, \bibinfo {author} {\bibfnamefont {Richard}\
  \bibnamefont {Cleve}}, \ and\ \bibinfo {author} {\bibfnamefont {Barry~C.}\
  \bibnamefont {Sanders}},\ }\bibfield  {title} {\enquote {\bibinfo {title}
  {Efficient quantum algorithms for simulating sparse hamiltonians},}\ }\href
  {\doibase 10.1007/s00220-006-0150-x} {\bibfield  {journal} {\bibinfo
  {journal} {Communications in Mathematical Physics}\ }\textbf {\bibinfo
  {volume} {270}},\ \bibinfo {pages} {359--371} (\bibinfo {year}
  {2006})}\BibitemShut {NoStop}%
\bibitem [{\citenamefont {Kjaergaard}\ \emph {et~al.}(2022)\citenamefont
  {Kjaergaard}, \citenamefont {Schwartz}, \citenamefont {Greene}, \citenamefont
  {Samach}, \citenamefont {Bengtsson}, \citenamefont {O'Keeffe}, \citenamefont
  {McNally}, \citenamefont {Braum\"uller}, \citenamefont {Kim}, \citenamefont
  {Krantz}, \citenamefont {Marvian}, \citenamefont {Melville}, \citenamefont
  {Niedzielski}, \citenamefont {Sung}, \citenamefont {Winik}, \citenamefont
  {Yoder}, \citenamefont {Rosenberg}, \citenamefont {Obenland}, \citenamefont
  {Lloyd}, \citenamefont {Orlando}, \citenamefont {Marvian}, \citenamefont
  {Gustavsson},\ and\ \citenamefont {Oliver}}]{LMR_experiment}%
  \BibitemOpen
  \bibfield  {author} {\bibinfo {author} {\bibfnamefont {M.}~\bibnamefont
  {Kjaergaard}}, \bibinfo {author} {\bibfnamefont {M.~E.}\ \bibnamefont
  {Schwartz}}, \bibinfo {author} {\bibfnamefont {A.}~\bibnamefont {Greene}},
  \bibinfo {author} {\bibfnamefont {G.~O.}\ \bibnamefont {Samach}}, \bibinfo
  {author} {\bibfnamefont {A.}~\bibnamefont {Bengtsson}}, \bibinfo {author}
  {\bibfnamefont {M.}~\bibnamefont {O'Keeffe}}, \bibinfo {author}
  {\bibfnamefont {C.~M.}\ \bibnamefont {McNally}}, \bibinfo {author}
  {\bibfnamefont {J.}~\bibnamefont {Braum\"uller}}, \bibinfo {author}
  {\bibfnamefont {D.~K.}\ \bibnamefont {Kim}}, \bibinfo {author} {\bibfnamefont
  {P.}~\bibnamefont {Krantz}}, \bibinfo {author} {\bibfnamefont
  {M.}~\bibnamefont {Marvian}}, \bibinfo {author} {\bibfnamefont
  {A.}~\bibnamefont {Melville}}, \bibinfo {author} {\bibfnamefont {B.~M.}\
  \bibnamefont {Niedzielski}}, \bibinfo {author} {\bibfnamefont
  {Y.}~\bibnamefont {Sung}}, \bibinfo {author} {\bibfnamefont {R.}~\bibnamefont
  {Winik}}, \bibinfo {author} {\bibfnamefont {J.}~\bibnamefont {Yoder}},
  \bibinfo {author} {\bibfnamefont {D.}~\bibnamefont {Rosenberg}}, \bibinfo
  {author} {\bibfnamefont {K.}~\bibnamefont {Obenland}}, \bibinfo {author}
  {\bibfnamefont {S.}~\bibnamefont {Lloyd}}, \bibinfo {author} {\bibfnamefont
  {T.~P.}\ \bibnamefont {Orlando}}, \bibinfo {author} {\bibfnamefont
  {I.}~\bibnamefont {Marvian}}, \bibinfo {author} {\bibfnamefont
  {S.}~\bibnamefont {Gustavsson}}, \ and\ \bibinfo {author} {\bibfnamefont
  {W.~D.}\ \bibnamefont {Oliver}},\ }\bibfield  {title} {\enquote {\bibinfo
  {title} {Demonstration of density matrix exponentiation using a
  superconducting quantum processor},}\ }\href {\doibase
  10.1103/PhysRevX.12.011005} {\bibfield  {journal} {\bibinfo  {journal} {Phys.
  Rev. X}\ }\textbf {\bibinfo {volume} {12}},\ \bibinfo {pages} {011005}
  (\bibinfo {year} {2022})}\BibitemShut {NoStop}%
\bibitem [{\citenamefont {Zhao}\ \emph {et~al.}(2019)\citenamefont {Zhao},
  \citenamefont {Pozas-Kerstjens}, \citenamefont {Rebentrost},\ and\
  \citenamefont {Wittek}}]{Zhao_2019}%
  \BibitemOpen
  \bibfield  {author} {\bibinfo {author} {\bibfnamefont {Zhikuan}\ \bibnamefont
  {Zhao}}, \bibinfo {author} {\bibfnamefont {Alejandro}\ \bibnamefont
  {Pozas-Kerstjens}}, \bibinfo {author} {\bibfnamefont {Patrick}\ \bibnamefont
  {Rebentrost}}, \ and\ \bibinfo {author} {\bibfnamefont {Peter}\ \bibnamefont
  {Wittek}},\ }\bibfield  {title} {\enquote {\bibinfo {title} {Bayesian deep
  learning on a quantum computer},}\ }\href {\doibase
  10.1007/s42484-019-00004-7} {\bibfield  {journal} {\bibinfo  {journal}
  {Quantum Machine Intelligence}\ }\textbf {\bibinfo {volume} {1}},\ \bibinfo
  {pages} {41--51} (\bibinfo {year} {2019})}\BibitemShut {NoStop}%
\bibitem [{\citenamefont {Schuld}\ \emph {et~al.}(2016)\citenamefont {Schuld},
  \citenamefont {Sinayskiy},\ and\ \citenamefont
  {Petruccione}}]{Schuld_regression}%
  \BibitemOpen
  \bibfield  {author} {\bibinfo {author} {\bibfnamefont {Maria}\ \bibnamefont
  {Schuld}}, \bibinfo {author} {\bibfnamefont {Ilya}\ \bibnamefont
  {Sinayskiy}}, \ and\ \bibinfo {author} {\bibfnamefont {Francesco}\
  \bibnamefont {Petruccione}},\ }\bibfield  {title} {\enquote {\bibinfo {title}
  {Prediction by linear regression on a quantum computer},}\ }\href {\doibase
  10.1103/physreva.94.022342} {\bibfield  {journal} {\bibinfo  {journal}
  {Physical Review A}\ }\textbf {\bibinfo {volume} {94}} (\bibinfo {year}
  {2016}),\ 10.1103/physreva.94.022342}\BibitemShut {NoStop}%
\bibitem [{\citenamefont {Cong}\ and\ \citenamefont
  {Duan}(2016)}]{discriminant}%
  \BibitemOpen
  \bibfield  {author} {\bibinfo {author} {\bibfnamefont {Iris}\ \bibnamefont
  {Cong}}\ and\ \bibinfo {author} {\bibfnamefont {Luming}\ \bibnamefont
  {Duan}},\ }\bibfield  {title} {\enquote {\bibinfo {title} {Quantum
  discriminant analysis for dimensionality reduction and classification},}\
  }\href {\doibase 10.1088/1367-2630/18/7/073011} {\bibfield  {journal}
  {\bibinfo  {journal} {New Journal of Physics}\ }\textbf {\bibinfo {volume}
  {18}},\ \bibinfo {pages} {073011} (\bibinfo {year} {2016})}\BibitemShut
  {NoStop}%
\bibitem [{\citenamefont {Liu}\ \emph {et~al.}(2022)\citenamefont {Liu},
  \citenamefont {Shen}, \citenamefont {Li}, \citenamefont {Duan},\ and\
  \citenamefont {Deng}}]{quantum_capsule}%
  \BibitemOpen
  \bibfield  {author} {\bibinfo {author} {\bibfnamefont {Zidu}\ \bibnamefont
  {Liu}}, \bibinfo {author} {\bibfnamefont {Pei-Xin}\ \bibnamefont {Shen}},
  \bibinfo {author} {\bibfnamefont {Weikang}\ \bibnamefont {Li}}, \bibinfo
  {author} {\bibfnamefont {L-M}\ \bibnamefont {Duan}}, \ and\ \bibinfo {author}
  {\bibfnamefont {Dong-Ling}\ \bibnamefont {Deng}},\ }\bibfield  {title}
  {\enquote {\bibinfo {title} {Quantum capsule networks},}\ }\href {\doibase
  10.1088/2058-9565/aca55d} {\bibfield  {journal} {\bibinfo  {journal} {Quantum
  Science and Technology}\ }\textbf {\bibinfo {volume} {8}},\ \bibinfo {pages}
  {015016} (\bibinfo {year} {2022})}\BibitemShut {NoStop}%
\bibitem [{\citenamefont {Wootters}\ and\ \citenamefont
  {Zurek}(1982)}]{no_clonning}%
  \BibitemOpen
  \bibfield  {author} {\bibinfo {author} {\bibfnamefont {W.~K.}\ \bibnamefont
  {Wootters}}\ and\ \bibinfo {author} {\bibfnamefont {W.~H.}\ \bibnamefont
  {Zurek}},\ }\bibfield  {title} {\enquote {\bibinfo {title} {A single quantum
  cannot be cloned},}\ }\href {\doibase 10.1038/299802a0} {\bibfield  {journal}
  {\bibinfo  {journal} {Nature}\ }\textbf {\bibinfo {volume} {299}},\ \bibinfo
  {pages} {802--803} (\bibinfo {year} {1982})}\BibitemShut {NoStop}%
\bibitem [{\citenamefont {Bu\ifmmode~\check{z}\else \v{z}\fi{}ek}\ and\
  \citenamefont {Hillery}(1996)}]{PhysRevA.54.1844}%
  \BibitemOpen
  \bibfield  {author} {\bibinfo {author} {\bibfnamefont {V.}~\bibnamefont
  {Bu\ifmmode~\check{z}\else \v{z}\fi{}ek}}\ and\ \bibinfo {author}
  {\bibfnamefont {M.}~\bibnamefont {Hillery}},\ }\bibfield  {title} {\enquote
  {\bibinfo {title} {Quantum copying: Beyond the no-cloning theorem},}\ }\href
  {\doibase 10.1103/PhysRevA.54.1844} {\bibfield  {journal} {\bibinfo
  {journal} {Phys. Rev. A}\ }\textbf {\bibinfo {volume} {54}},\ \bibinfo
  {pages} {1844--1852} (\bibinfo {year} {1996})}\BibitemShut {NoStop}%
\bibitem [{\citenamefont {Wang}\ \emph {et~al.}(2011)\citenamefont {Wang},
  \citenamefont {Shi}, \citenamefont {Xiong}, \citenamefont {Jing},
  \citenamefont {Ren}, \citenamefont {Mu},\ and\ \citenamefont
  {Fan}}]{Wang_2011}%
  \BibitemOpen
  \bibfield  {author} {\bibinfo {author} {\bibfnamefont {Yi-Nan}\ \bibnamefont
  {Wang}}, \bibinfo {author} {\bibfnamefont {Han-Duo}\ \bibnamefont {Shi}},
  \bibinfo {author} {\bibfnamefont {Zhao-Xi}\ \bibnamefont {Xiong}}, \bibinfo
  {author} {\bibfnamefont {Li}~\bibnamefont {Jing}}, \bibinfo {author}
  {\bibfnamefont {Xi-Jun}\ \bibnamefont {Ren}}, \bibinfo {author}
  {\bibfnamefont {Liang-Zhu}\ \bibnamefont {Mu}}, \ and\ \bibinfo {author}
  {\bibfnamefont {Heng}\ \bibnamefont {Fan}},\ }\bibfield  {title} {\enquote
  {\bibinfo {title} {Unified universal quantum cloning machine and
  fidelities},}\ }\href {\doibase 10.1103/physreva.84.034302} {\bibfield
  {journal} {\bibinfo  {journal} {Physical Review A}\ }\textbf {\bibinfo
  {volume} {84}} (\bibinfo {year} {2011}),\
  10.1103/physreva.84.034302}\BibitemShut {NoStop}%
\bibitem [{\citenamefont {Alvarez-Rodriguez}\ \emph {et~al.}(2014)\citenamefont
  {Alvarez-Rodriguez}, \citenamefont {Sanz}, \citenamefont {Lamata},\ and\
  \citenamefont {Solano}}]{Alvarez-Rodriguez_2014}%
  \BibitemOpen
  \bibfield  {author} {\bibinfo {author} {\bibfnamefont {U.}~\bibnamefont
  {Alvarez-Rodriguez}}, \bibinfo {author} {\bibfnamefont {M.}~\bibnamefont
  {Sanz}}, \bibinfo {author} {\bibfnamefont {L.}~\bibnamefont {Lamata}}, \ and\
  \bibinfo {author} {\bibfnamefont {E.}~\bibnamefont {Solano}},\ }\bibfield
  {title} {\enquote {\bibinfo {title} {Biomimetic cloning of quantum
  observables},}\ }\href {\doibase 10.1038/srep04910} {\bibfield  {journal}
  {\bibinfo  {journal} {Scientific Reports}\ }\textbf {\bibinfo {volume} {4}},\
  \bibinfo {pages} {4910} (\bibinfo {year} {2014})}\BibitemShut {NoStop}%
\bibitem [{\citenamefont {Holmes}\ \emph {et~al.}(2023)\citenamefont {Holmes},
  \citenamefont {Coble}, \citenamefont {Sornborger},\ and\ \citenamefont
  {Suba\ifmmode \mbox{\c{s}}\else \c{s}\fi{}\ifmmode \imath \else~\i
  \fi{}}}]{Zoe_holmes}%
  \BibitemOpen
  \bibfield  {author} {\bibinfo {author} {\bibfnamefont {Zo\"e}\ \bibnamefont
  {Holmes}}, \bibinfo {author} {\bibfnamefont {Nolan~J.}\ \bibnamefont
  {Coble}}, \bibinfo {author} {\bibfnamefont {Andrew~T.}\ \bibnamefont
  {Sornborger}}, \ and\ \bibinfo {author} {\bibfnamefont {Yi\ifmmode
  \breve{g}\else~\u{g}\fi{}it}\ \bibnamefont {Suba\ifmmode \mbox{\c{s}}\else
  \c{s}\fi{}\ifmmode \imath \else~\i \fi{}}},\ }\bibfield  {title} {\enquote
  {\bibinfo {title} {Nonlinear transformations in quantum computation},}\
  }\href {\doibase 10.1103/PhysRevResearch.5.013105} {\bibfield  {journal}
  {\bibinfo  {journal} {Phys. Rev. Res.}\ }\textbf {\bibinfo {volume} {5}},\
  \bibinfo {pages} {013105} (\bibinfo {year} {2023})}\BibitemShut {NoStop}%
\bibitem [{\citenamefont {Huggins}\ \emph {et~al.}(2021)\citenamefont
  {Huggins}, \citenamefont {McArdle}, \citenamefont {O'Brien}, \citenamefont
  {Lee}, \citenamefont {Rubin}, \citenamefont {Boixo}, \citenamefont {Whaley},
  \citenamefont {Babbush},\ and\ \citenamefont {McClean}}]{Huggins_2021}%
  \BibitemOpen
  \bibfield  {author} {\bibinfo {author} {\bibfnamefont {William~J.}\
  \bibnamefont {Huggins}}, \bibinfo {author} {\bibfnamefont {Sam}\ \bibnamefont
  {McArdle}}, \bibinfo {author} {\bibfnamefont {Thomas~E.}\ \bibnamefont
  {O'Brien}}, \bibinfo {author} {\bibfnamefont {Joonho}\ \bibnamefont {Lee}},
  \bibinfo {author} {\bibfnamefont {Nicholas~C.}\ \bibnamefont {Rubin}},
  \bibinfo {author} {\bibfnamefont {Sergio}\ \bibnamefont {Boixo}}, \bibinfo
  {author} {\bibfnamefont {K.~Birgitta}\ \bibnamefont {Whaley}}, \bibinfo
  {author} {\bibfnamefont {Ryan}\ \bibnamefont {Babbush}}, \ and\ \bibinfo
  {author} {\bibfnamefont {Jarrod~R.}\ \bibnamefont {McClean}},\ }\bibfield
  {title} {\enquote {\bibinfo {title} {Virtual distillation for quantum error
  mitigation},}\ }\href {\doibase 10.1103/physrevx.11.041036} {\bibfield
  {journal} {\bibinfo  {journal} {Physical Review X}\ }\textbf {\bibinfo
  {volume} {11}} (\bibinfo {year} {2021}),\
  10.1103/physrevx.11.041036}\BibitemShut {NoStop}%
\bibitem [{\citenamefont {Huang}\ \emph {et~al.}(2022)\citenamefont {Huang},
  \citenamefont {Broughton}, \citenamefont {Cotler}, \citenamefont {Chen},
  \citenamefont {Li}, \citenamefont {Mohseni}, \citenamefont {Neven},
  \citenamefont {Babbush}, \citenamefont {Kueng}, \citenamefont {Preskill},\
  and\ \citenamefont {McClean}}]{Huang_2022}%
  \BibitemOpen
  \bibfield  {author} {\bibinfo {author} {\bibfnamefont {Hsin-Yuan}\
  \bibnamefont {Huang}}, \bibinfo {author} {\bibfnamefont {Michael}\
  \bibnamefont {Broughton}}, \bibinfo {author} {\bibfnamefont {Jordan}\
  \bibnamefont {Cotler}}, \bibinfo {author} {\bibfnamefont {Sitan}\
  \bibnamefont {Chen}}, \bibinfo {author} {\bibfnamefont {Jerry}\ \bibnamefont
  {Li}}, \bibinfo {author} {\bibfnamefont {Masoud}\ \bibnamefont {Mohseni}},
  \bibinfo {author} {\bibfnamefont {Hartmut}\ \bibnamefont {Neven}}, \bibinfo
  {author} {\bibfnamefont {Ryan}\ \bibnamefont {Babbush}}, \bibinfo {author}
  {\bibfnamefont {Richard}\ \bibnamefont {Kueng}}, \bibinfo {author}
  {\bibfnamefont {John}\ \bibnamefont {Preskill}}, \ and\ \bibinfo {author}
  {\bibfnamefont {Jarrod~R.}\ \bibnamefont {McClean}},\ }\bibfield  {title}
  {\enquote {\bibinfo {title} {Quantum advantage in learning from
  experiments},}\ }\href {\doibase 10.1126/science.abn7293} {\bibfield
  {journal} {\bibinfo  {journal} {Science}\ }\textbf {\bibinfo {volume}
  {376}},\ \bibinfo {pages} {1182--1186} (\bibinfo {year} {2022})}\BibitemShut
  {NoStop}%
\bibitem [{\citenamefont {Go}\ \emph {et~al.}(2024)\citenamefont {Go},
  \citenamefont {Kwon}, \citenamefont {Park}, \citenamefont {Patel},\ and\
  \citenamefont {Wilde}}]{bound_dme}%
  \BibitemOpen
  \bibfield  {author} {\bibinfo {author} {\bibfnamefont {Byeongseon}\
  \bibnamefont {Go}}, \bibinfo {author} {\bibfnamefont {Hyukjoon}\ \bibnamefont
  {Kwon}}, \bibinfo {author} {\bibfnamefont {Siheon}\ \bibnamefont {Park}},
  \bibinfo {author} {\bibfnamefont {Dhrumil}\ \bibnamefont {Patel}}, \ and\
  \bibinfo {author} {\bibfnamefont {Mark~M.}\ \bibnamefont {Wilde}},\ }\href
  {https://arxiv.org/abs/2412.02134} {\enquote {\bibinfo {title} {Density
  matrix exponentiation and sample-based hamiltonian simulation: Non-asymptotic
  analysis of sample complexity},}\ } (\bibinfo {year} {2024}),\ \Eprint
  {http://arxiv.org/abs/2412.02134} {arXiv:2412.02134 [quant-ph]} \BibitemShut
  {NoStop}%
\bibitem [{\citenamefont {Welch}\ \emph {et~al.}(2014)\citenamefont {Welch},
  \citenamefont {Greenbaum}, \citenamefont {Mostame},\ and\ \citenamefont
  {Aspuru-Guzik}}]{Welch_2014}%
  \BibitemOpen
  \bibfield  {author} {\bibinfo {author} {\bibfnamefont {Jonathan}\
  \bibnamefont {Welch}}, \bibinfo {author} {\bibfnamefont {Daniel}\
  \bibnamefont {Greenbaum}}, \bibinfo {author} {\bibfnamefont {Sarah}\
  \bibnamefont {Mostame}}, \ and\ \bibinfo {author} {\bibfnamefont {Alan}\
  \bibnamefont {Aspuru-Guzik}},\ }\bibfield  {title} {\enquote {\bibinfo
  {title} {Efficient quantum circuits for diagonal unitaries without
  ancillas},}\ }\href {\doibase 10.1088/1367-2630/16/3/033040} {\bibfield
  {journal} {\bibinfo  {journal} {New Journal of Physics}\ }\textbf {\bibinfo
  {volume} {16}},\ \bibinfo {pages} {033040} (\bibinfo {year}
  {2014})}\BibitemShut {NoStop}%
\bibitem [{\citenamefont {Takahashi}(2009)}]{takahashi2009quantum}%
  \BibitemOpen
  \bibfield  {author} {\bibinfo {author} {\bibfnamefont {Yasuhiro}\
  \bibnamefont {Takahashi}},\ }\bibfield  {title} {\enquote {\bibinfo {title}
  {Quantum arithmetic circuits: A survey},}\ }\href
  {https://api.semanticscholar.org/CorpusID:6749815} {\bibfield  {journal}
  {\bibinfo  {journal} {IEICE Trans. Fundam. Electron. Commun. Comput. Sci.}\
  }\textbf {\bibinfo {volume} {92-A}},\ \bibinfo {pages} {1276--1283} (\bibinfo
  {year} {2009})}\BibitemShut {NoStop}%
\bibitem [{\citenamefont {Pucha{\l}a}\ \emph {et~al.}(2016)\citenamefont
  {Pucha{\l}a}, \citenamefont {Pawela},\ and\ \citenamefont {{\.{Z}
  }yczkowski}}]{Pucha_a_2016}%
  \BibitemOpen
  \bibfield  {author} {\bibinfo {author} {\bibfnamefont {Zbigniew}\
  \bibnamefont {Pucha{\l}a}}, \bibinfo {author} {\bibfnamefont {{\L}ukasz}\
  \bibnamefont {Pawela}}, \ and\ \bibinfo {author} {\bibfnamefont {Karol}\
  \bibnamefont {{\.{Z} }yczkowski}},\ }\bibfield  {title} {\enquote {\bibinfo
  {title} {Distinguishability of generic quantum states},}\ }\href {\doibase
  10.1103/physreva.93.062112} {\bibfield  {journal} {\bibinfo  {journal}
  {Physical Review A}\ }\textbf {\bibinfo {volume} {93}} (\bibinfo {year}
  {2016}),\ 10.1103/physreva.93.062112}\BibitemShut {NoStop}%
\end{thebibliography}%
\end{document}